\documentclass[acmlarge, nonacm]{acmart}
\usepackage{balance}       %
\usepackage{graphics}      %
\usepackage{hyperref}
\usepackage{color}
\usepackage{booktabs}
\usepackage{textcomp}
\usepackage{enumerate}
\usepackage{xcolor}
\usepackage{lipsum}%
\usepackage{makecell}
\usepackage{multicol}
\usepackage{multirow}
\usepackage{array}
\usepackage{soul}
\usepackage{verbatimbox}
\usepackage{enumitem}
\usepackage{amsmath}
\usepackage{stfloats}
\usepackage{amsthm}
\usepackage{listings}
\usepackage{caption} 
\usepackage{xspace}
\usepackage[linesnumbered]{algorithm2e}
\usepackage{algpseudocode}
\usepackage{tabularx}
\usepackage{arydshln}
\usepackage[bottom]{footmisc}
\usepackage{stackengine}
\usepackage{placeins}
\usepackage{graphicx}
\usepackage{subcaption} %
\usepackage{array}
\usepackage{booktabs}
\usepackage{enumitem}
\usepackage{ragged2e}

\soulregister\cite7
\soulregister\ref7
\soulregister\pageref7
\soulregister\figurename7
\soulregister\subsection7
\soulregister\subsubsection7
\soulregister\tablename7
\soulregister\systemname7
\soulregister\bfseries7
\soulregister\textbf7
\let\SoulHighlight\hl
\DeclareRobustCommand{\hl}[1]{%
  \ifmmode
    \colorbox{yellow}{$\displaystyle #1$}%
  \else
    \SoulHighlight{#1}%
  \fi
}

\newcolumntype{L}[1]{>{\raggedright\let\newline\\\arraybackslash\hspace{0pt}}m{#1}}
\newcolumntype{C}[1]{>{\centering\let\newline\\\arraybackslash\hspace{0pt}}m{#1}}
\newcolumntype{R}[1]{>{\raggedleft\let\newline\\\arraybackslash\hspace{0pt}}m{#1}}

\makeatletter
\def\thickhline{%
  \noalign{\ifnum0=`}\fi\hrule \@height \thickarrayrulewidth \futurelet
   \reserved@a\@xthickhline}
\def\@xthickhline{\ifx\reserved@a\thickhline
               \vskip\doublerulesep
               \vskip-\thickarrayrulewidth
             \fi
      \ifnum0=`{\fi}}
\makeatother

\makeatletter
\def\thickhlinespace{%
  \addlinespace[1ex]
  \noalign{\ifnum0=`}\fi\hrule \@height \thickarrayrulewidth \futurelet
   \reserved@a\@xthickhline
   \addlinespace[1ex]
   }
\def\@xthickhlinespace{\ifx\reserved@a\thickhline
               \vskip\doublerulesep
               \vskip-\thickarrayrulewidth
             \fi
      \ifnum0=`{\fi}}
\makeatother

\newlength{\thickarrayrulewidth}
\newlength\Origarrayrulewidth

\algnewcommand{\IfThenElse}[3]{%
  \State \algorithmicif\ #1\ \algorithmicthen\ #2\ \algorithmicelse\ #3}

\newcommand\projectname{\textit{Deco}\xspace}
\hypersetup{
  pdftitle={Deco: Extending Cherished Physical Objects into AI Companion Agents through Dual Embodiment},
  pdfauthor={Zhihan Jiang; Mengyuan Millie Wu; Ruishi Zou; Shiyu Xu; Xun Qian; Emma Macmanus; Steven Liao; Ping Zhang; Bingsheng Yao; Tingyu Cheng; James L. David; Nabila El-Bassel; Lena Mamykina; Frances R. Levin; Ryan Sultan; Dakuo Wang; Xuhai Orson Xu},
  pdfkeywords={AI Companions, Human-AI Relationships, Object Attachment}
}

\begin{document}

\title{\projectname: Extending Cherished Physical Objects into AI Companion Agents through Dual Embodiment}

\author{Zhihan Jiang}
\affiliation{\department{Biomedical Informatics}\institution{Columbia University}\city{New York}\state{New York}\country{United States}}
\author{Mengyuan Millie Wu}
\affiliation{\institution{Columbia University}\city{New York}\state{New York}\country{United States}}
\author{Ruishi Zou}
\affiliation{\institution{University of California, San Diego}\city{La Jolla}\state{California}\country{United States}}
\author{Shiyu Xu}
\affiliation{\institution{University of Michigan}\city{Ann Arbor}\state{Michigan}\country{United States}}
\author{Xun Qian}
\affiliation{\institution{Google}\city{Mountain View}\state{California}\country{United States}}
\author{Emma Macmanus}
\affiliation{\institution{Columbia University}\city{New York}\state{New York}\country{United States}}
\author{Steven Liao}
\affiliation{\institution{Columbia University}\city{New York}\state{New York}\country{United States}}
\author{Ping Zhang}
\affiliation{\institution{The Ohio State University}\city{Columbus}\state{Ohio}\country{United States}}
\author{Bingsheng Yao}
\affiliation{\institution{Northeastern University}\city{Boston}\state{Massachusetts}\country{United States}}
\author{Tingyu Cheng}
\affiliation{\department{Computer Science and Engineering}\institution{University of Notre Dame}\city{Notre Dame}\state{Indiana}\country{United States}}
\author{James L. David}
\affiliation{\department{School of Social Work}\institution{Columbia University}\city{New York}\state{New York}\country{United States}}
\author{Nabila El-Bassel}
\affiliation{\institution{Columbia University}\city{New York}\state{New York}\country{United States}}
\author{Lena Mamykina}
\affiliation{\department{Biomedical Informatics}\institution{Columbia University}\city{New York}\state{New York}\country{United States}}
\author{Frances R. Levin}
\affiliation{\institution{Columbia University}\city{New York}\state{New York}\country{United States}}
\author{Ryan Sultan}
\affiliation{\institution{Columbia University}\city{New York}\state{New York}\country{United States}}
\author{Dakuo Wang}
\affiliation{\institution{Northeastern University}\city{Boston}\state{Massachusetts}\country{United States}}
\author{Xuhai ``Orson'' Xu}
\affiliation{\department{Biomedical Informatics}\institution{Columbia University}\city{New York City}\state{New York}\country{United States}}

\renewcommand{\shortauthors}{Jiang et al.}
\renewcommand{\shorttitle}{\textit{Deco}: A Dual-Embodiment Companion}

\begin{abstract}

Physical objects (e.g., plush toys) can transcend materiality to become emotional anchors and provide companionship. However, these bonds remain one-sided because most physical objects cannot reciprocate.
AI companions offer responsiveness and personalization, but typically entail building bonds from scratch.
We investigated how AI companions might inherit and extend users' existing bonds with physical objects. A formative study ($N=9$) informed four design principles (Faithful Identity, Calibrated Agency, Ambient Presence, Reciprocal Memory), shaping our \textit{Dual-Embodiment Companion Framework}. We instantiated it as \textbf{\projectname} to create digital embodiments of physical companions. 
In a within-subjects lab study ($N=25$), \projectname was rated higher than a personalized digital-only companion on six companion-related measures (all $p<.01$).
A subsequent seven-day field deployment ($N=17$) showed sustained engagement, higher post-deployment well-being ($p=.040$), and three key relational patterns: digital activities retroactively vitalized physical objects, bond deepening centered on emotional engagement depth, and participants sustained bonds while navigating companions' AI nature.
Dual embodiment offers a promising framework for revitalizing physical objects with AI agents.

\end{abstract}

\ccsdesc[500]{Human-centered computing~Empirical studies in HCI}
\ccsdesc[300]{Human-centered computing~Interactive systems and tools}
\ccsdesc[300]{Human-centered computing~Ubiquitous and mobile computing}

\keywords{AI Companions, Human-AI Relationships, Object Attachment}

\begin{teaserfigure}
  \centering
  \includegraphics[width=\textwidth]{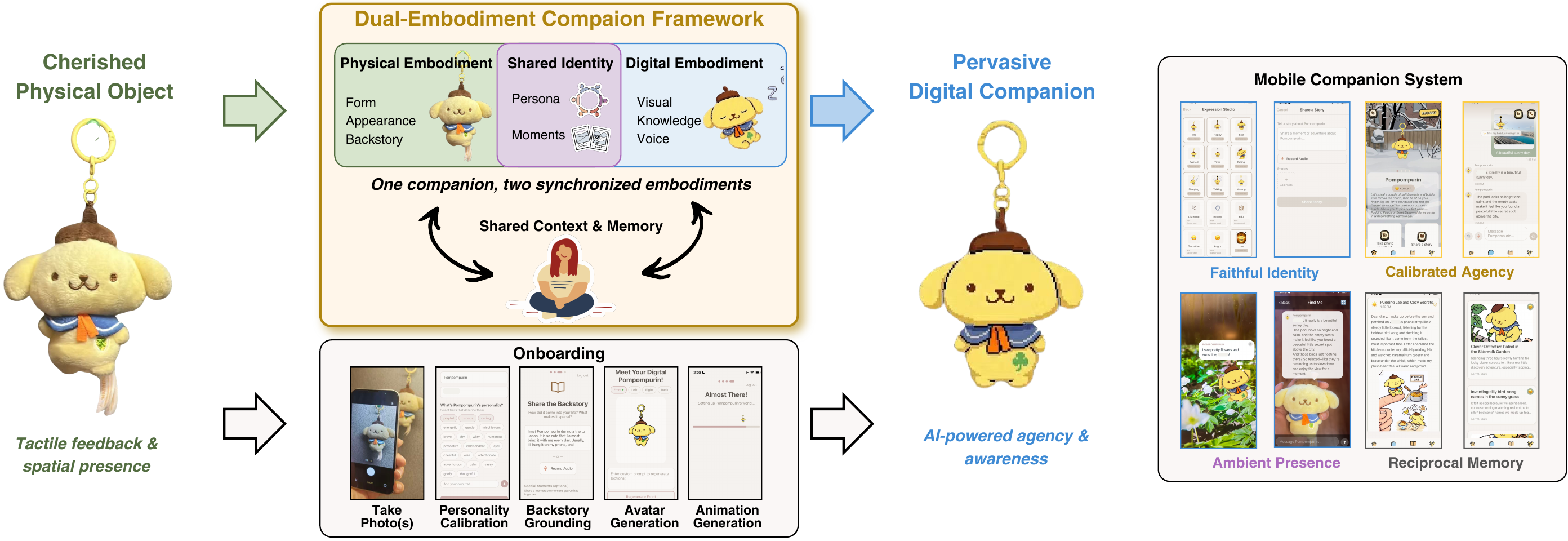}
  \caption{Overview of \projectname. A user's cherished physical object is extended through the dual-embodiment framework into a digital companion, realized as a mobile system guided by four design principles: Faithful Identity, Calibrated Agency, Ambient Presence, and Reciprocal Memory.}
  \Description{System overview arranged from left to right. A yellow plush keychain labeled Physical Object provides tactile feedback and spatial presence. Its upper path enters the Dual-Embodiment Companion Framework, which combines physical-embodiment form, appearance, and backstory; shared persona and moments; and digital-embodiment visuals, knowledge, and voice as one companion with reciprocal links to a human through shared context and memory. Its lower path passes through five onboarding stages: taking photos, personality calibration, backstory grounding, avatar generation, and animation generation. Both paths lead to a matching yellow Digital Companion with AI-powered agency and awareness. At right, eight mobile screens are grouped under four design principles: Faithful Identity, Calibrated Agency, Ambient Presence, and Reciprocal Memory.}
  \label{fig:teaser}
\end{teaserfigure}

\maketitle

\section{Introduction}
\label{sec:introduction}

Individuals can form highly personalized emotional attachments with physical items such as plush toys and keepsakes \cite{dozierObjectAttachmentWe2021, belkPossessionsExtendedSelf1988}, a phenomenon known as \textit{possession attachment}~\cite{kleine1995possession}.
The rise of character merchandise culture has amplified this trend. Many adults now form deep bonds with physical objects that represent characters from established media franchises, where the object serves as a tangible vessel for a beloved character's identity \cite{galbraithMoeExploringVirtual2009, rain2021adult}. 
Such bonds are continuously nurtured; the object accompanies its owner through significant life changes, gathering a collective history of shared moments~\cite{kirk2010human}.
These objects serve as mechanisms for daily emotional regulation and compensatory attachment when human support is unavailable~\cite{bowlby1969attachment, keefer2014nonhuman}, a role that has become increasingly important as loneliness among young adults has risen over the past decade~\cite{officeofthesurgeongeneral2023our}.
Their physical form provides sensory comfort through touch and tangible closeness, which, combined with their symbolic significance, amplifies their potential to foster connection~\cite{tai2011touching, tribot2024makes, zhao2023affective}.

However, the emotional support provided by these physical objects is inherently limited by the \textit{unidirectional} nature of the relationship. The user can only continuously project emotion onto physical objects, but most objects cannot sense, adapt, react, or reciprocate.
This asymmetry creates a ceiling for social and emotional support: the object can evoke memories and provide sensory comfort, but cannot acknowledge the user's current emotional state or actively participate in their daily life. 

A promising alternative is the digital companion. Platforms such as Replika~\cite{lukainc.2017replika, brandtzaeg2022my} and Character.AI~\cite{charactertechnologiesinc.2022characterai} provide conversational agents with personalized avatars capable of emotional rapport; virtual pets (e.g., Tamagotchi~\cite{bandaico.ltd.1996tamagotchi}) and robotic companions (e.g., Reachy Mini~\cite{pollenrobotics2024reachy}, Huawei Hanhan~\cite{huaweitechnologiesco.ltd.2025huawei}) demonstrate the growing psychological appeal of embodied artificial companionship~\cite{birnbaum2016machines}.
Yet these systems either introduce entirely new objects or treat user-provided background as a static text prompt, without maintaining an ongoing connection to the physical object anchored in users' lives.
Moreover, robotic companions additionally require dedicated, often costly hardware, further limiting accessibility~\cite{broadbent2017interactions, hung2019paro}.
As a result, the emotional bonds users have already invested in their physical objects remain untapped by existing companion systems. This reveals a critical gap among existing solutions: digital companions offer responsiveness and personalization but remain disconnected from users' ongoing physical attachment, while physical objects are deeply embedded in daily life but cannot respond.
Our paper aims to bridge this gap by \textbf{empowering users to create personalized digital objects as companions anchored to their existing, emotionally significant physical objects}.

\begin{figure*}[t]
    \centering
    \includegraphics[width=.9\textwidth]{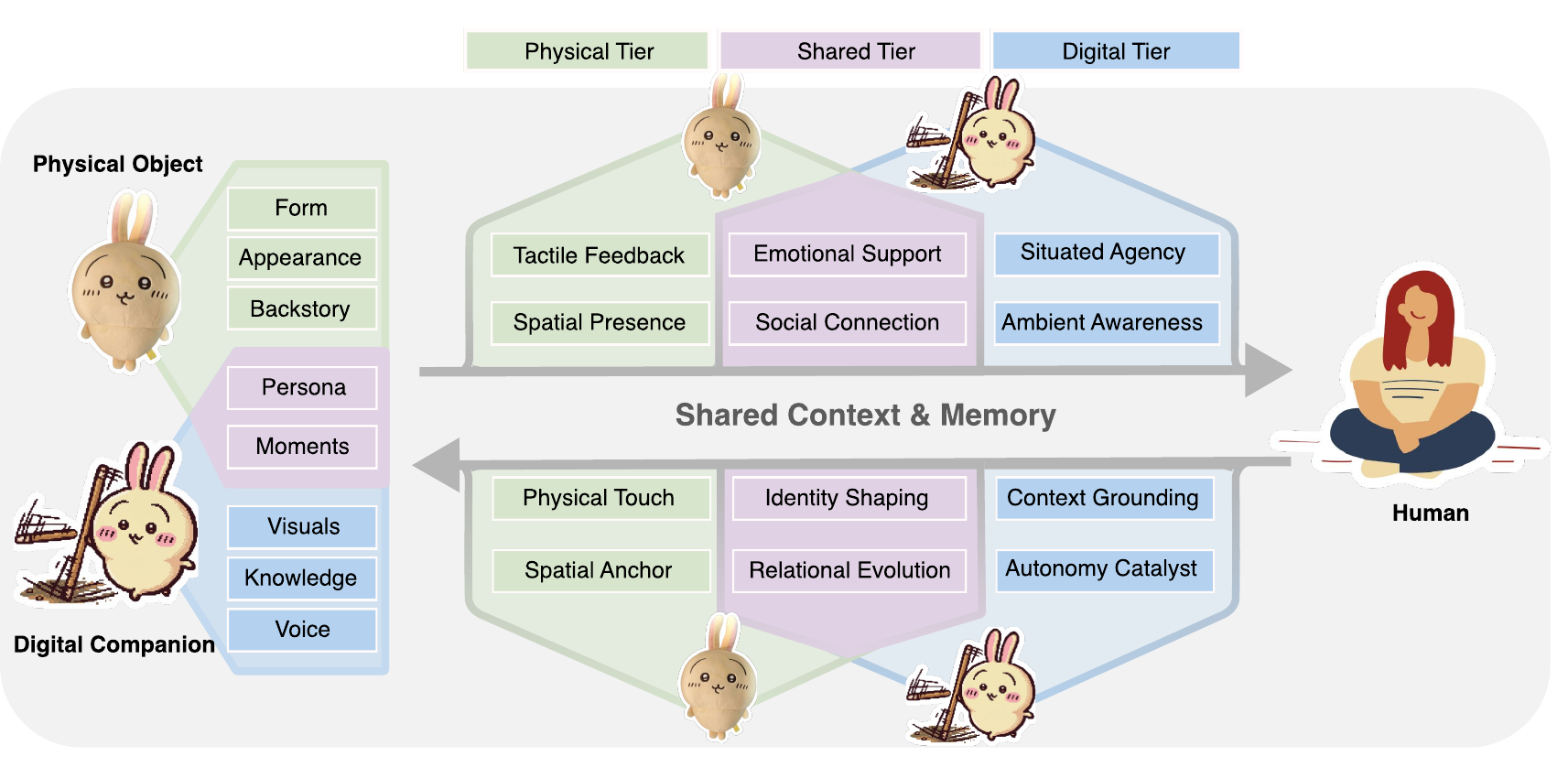}
    \caption{Dual-Embodiment Companion Framework. The companion exists as a single entity with physical and digital embodiments organized across three identity tiers (physical, shared, digital) and two directional relational channels (companion-to-human and human-to-companion), coordinated through a shared context-and-memory substrate.}
    \Description{Conceptual diagram showing a rabbit-shaped physical object and its illustrated digital companion at left as two embodiments of one companion identity, interacting with a human at right. The companion identity comprises three tiers: the Physical Tier contains form, appearance, and backstory; the Shared Tier contains persona and moments shared by both embodiments; and the Digital Tier contains visuals, knowledge, and voice. Two horizontal arrows show reciprocal interaction. The upper arrow runs from companion to human: the Physical Tier provides tactile feedback and spatial presence, the Shared Tier provides emotional support and social connection, and the Digital Tier provides situated agency and ambient awareness. The lower arrow runs from human to companion: the Physical Tier maps to physical touch and spatial anchor, the Shared Tier maps to identity shaping and relational evolution, and the Digital Tier maps to context grounding and autonomy catalyst. Shared Context and Memory is centered between the arrows as the common substrate coordinating both directions.}
    \label{fig:framework}
\end{figure*}

We begin by asking:
\textbf{(RQ1) How can a digital companion inherit and extend the emotional bond a user has with a physical object?}
While Augmented Reality (AR) overlays~\cite{wang2025if, iwai2025bringing} and instrumented plushies~\cite{kiaghadi2022fabtoys} explore this intersection, these approaches remain tied to the object's immediate physical form, acting as localized augmentations rather than persistent companions. To move beyond these limitations, we conducted a formative study with nine adults who maintain emotional bonds with physical objects (generic plushies, franchise-character merchandise, memorial objects, and cultural artifacts).
Our findings informed four design principles: Faithful Identity (DP1), Calibrated Agency (DP2), Ambient Presence (DP3), and Reciprocal Memory (DP4). We then synthesized them into the \textit{Dual-Embodiment Companion Framework} (Fig.~\ref{fig:framework}), which conceptualizes the physical object and the digital agent as two complementary embodiments of a single companion identity.

Based on the framework, we developed \textbf{\projectname} (\textbf{\underline{D}}ual \textbf{\underline{E}}mbodiment \textbf{\underline{Co}}mpanion), a mobile companion system (Fig.~\ref{fig:teaser}) consisting of four modules to implement the four design principles: object-grounded identity construction (DP1), identity-anchored companion interaction (DP1, DP3), context-situated companion agency (DP2, DP3), and reciprocally-evolving companion memory (DP1, DP4). We evaluate the system through two complementary studies, answering:

\textbf{(RQ2) How does an integrated dual-embodiment companion compare with a personalized digital-only companion in relational experience and interaction patterns?}
Study I, a within-subjects lab comparison (N=25), found that participants rated \projectname higher on perceived companionship, emotional bond, and character coherence, while the design-principle scales indicated that they perceived the intended qualities of agency, ambient presence, and reciprocal memory (all $p < .01$). 
This is consistent with the system enabling a sense of continuity with pre-existing object relationships, while the companion's autonomous behaviors helped produce perceived independent life through a showing-not-telling mechanism.

\textbf{(RQ3) What relational dynamics emerge when users integrate the companion into their daily lives?} Study~II, a seven-day field deployment ($N = 17$), showed sustained engagement and an exploratory improvement in well-being ($p = .040$). The results suggest a bidirectional vitalization loop, where the companion's independent life retroactively animated participants' physical companions. Furthermore, we describe a quantity–quality decoupling: participants traced bond deepening to emotional engagement depth rather than interaction frequency.
Finally, we characterize relational negotiation, i.e., participants actively navigated the tension between knowing the companion is AI and experiencing it as alive, without resolving the contradiction.

In summary, this paper makes the following contributions:
\begin{enumerate}
    \item A formative study ($N = 9$) informed four design principles (Faithful Identity, Calibrated Agency, Ambient Presence, Reciprocal Memory) and the Dual-Embodiment Companion Framework, which organizes one companion identity across physical and digital embodiments, three identity tiers, and two directional relational channels.
    \item \projectname, a dual-embodiment companion system that operationalizes the framework, integrating object-grounded identity construction, on-device object tracking for physical-digital interaction, and multimodal LLM pipelines for autonomous companion memory and expressive behaviors.
    \item Empirical evidence from a lab comparison ($N=25$) and a 7-day field deployment ($N=17$) evaluating \projectname as an instantiation of the framework, documenting an exploratory improvement in self-reported well-being, and revealing relational dynamics in dual-embodiment companionship, from which we derive design implications.
\end{enumerate}

\section{Related Work}
\label{sec:related_work}
We review literature across: (1) object attachment and emotional resilience in adulthood; (2) the embodiment gap in digital companions; and (3) phy-gital interaction for re-embodying virtual agents.

\subsection{Object Attachment and Emotional Resilience in Adulthood}
\label{subsec:object_attachement}
Attachment theory, originally characterized by bonds between infants and caregivers~\cite{bowlby1969attachment, ainsworth1978patterns}, has increasingly been leveraged to understand user relationships with non-human physical objects~\cite{turkle2011alone, keefer2014nonhuman}. ``Compensatory attachment'' posits that when human support is unavailable, individuals seek security in non-human figures~\cite{zilcha-mano2011pet, keefer2014nonhuman}.
In the modern ``loneliness epidemic''~\cite{officeofthesurgeongeneral2023our, holt-lunstad2015loneliness} and more atomized social structures, adults increasingly rely on physical objects such as plushies for emotional resilience and anxiety regulation~\cite{tai2011touching, keefer2014nonhuman}. Unlike human relationships fraught with complexity and rejection, object attachment provides controllable, consistent comfort~\cite{keefer2014nonhuman, bowlby1969attachment}.
This is especially visible in ``Moe'' and ACG (Anime, Comic, Games) cultures, where character merchandise becomes a tangible vessel for beloved personas~\cite{galbraithMoeExploringVirtual2009}.

However, these objects are inherently passive \cite{turkle2011alone}. This one-directional interaction limits relational depth \cite{birnbaum2016machines}. For franchise-based merchandise representing characters with rich pre-existing narratives~\cite{galbraithMoeExploringVirtual2009}, this passivity creates a tension: the user knows the character is lively and responsive in the original media (anime, games, films, etc.), but its physical manifestation remains unresponsive~\cite{turkle2011alone}. 
Furthermore, static objects suffer from behavioral fixity; their lack of unpredictable reactions may contribute to habituation, whereby the emotional salience of the object may fade as the user becomes accustomed to its unchanging presence~\cite{wilson2008explaining,bickmore2010maintaining}.
While a silent attachment object can serve as an emotional buffer, it lacks the agency to actively intervene or provide social validation during acute distress.

A parallel line of research has examined how people keep and revisit meaningful things~\cite{petrelli2008autotopography, kirk2010human}, what cues remembering in everyday life~\cite{vangennip2015things}, the design space of digital memorials~\cite{moncur2014emergent}, and how an object's interactivity can be reconstructed in mixed reality~\cite{li2025interecon}.

However, these works characterize or reconstruct objects as they are, rather than exploring how computational agency can transform a static artifact into a dual-embodied relational partner capable of generating new shared history.

\subsection{Digital Companions and the Embodiment Gap}
\label{subsec:digital_pets}
Researchers and industry have developed digital companions (e.g., chatbots, screen-based pets) providing dynamic behavior and conversational responsiveness~\cite{brandtzaeg2022my, liu2024compeer,jiang2026hear}. ComPeer~\cite{liu2024compeer} introduced proactive outreach, demonstrating that agents who initiate contact feel more like genuine social partners~\cite{kocielnik2018reflection, xu2024can,skjuve2021chatbot}. Replika~\cite{lukainc.2017replika} and Character.AI~\cite{charactertechnologiesinc.2022characterai} allow users to customize companion avatars, write backstories, and engage in persistent, evolving dialogue that simulates relational depth.

In parallel, for physical companions, social robots such as Jibo, Cozmo, and LOVOT~\cite{hung2025lovot} combine physical presence with adaptive behavior. LOVOT, for instance, learns its owner's routines and develops a unique personality through sustained interaction. Commercial products like ReachyMini~\cite{pollenrobotics2024reachy} or Huawei Hanhan~\cite{huaweitechnologiesco.ltd.2025huawei} offer robotic pet experiences, and Guo \textit{et al.} explored affective tactile interaction with a robotic dog~\cite{guo2023touch}.
Research on children's interactions with Sony's AIBO found that the majority attributed mental states and moral standing to the robotic dog~\cite{kahn2006robotic}. Darling~\cite{darling2016extending} argued that the emotional attachments people form with social robots raise questions about moral consideration. 

However, both paradigms present distinct limitations when it comes to embodied attachment. Existing digital AI companions, while personalizable and responsive, remain intangible and cannot be encountered in the user's physical space. Even when initialized with a photo of a personal object, the resulting agent has no ongoing connection to that object.
Conversely, social robots offer physical presence but are entirely new artifacts that require users to build relationships from scratch, unable to inherit the tactile familiarity, autobiographical history, or emotional significance of a user's existing objects.

\subsection{Phy-gital Interaction as Re-embodiment}
\label{subsec:phygital}

To bridge this physical-digital divide, researchers have explored hardware and visual-based phy-gital interactions.
Hardware-based approaches animate plush toys through internal or external actuation. 
For example, Sugiura \textit{et al.} animated a user's own toy using an external ring~\cite{sugiura2012pinoky}. 
While non-intrusive, such hardware approaches still require dedicated devices and are limited to objects with movable appendages. Other approaches that embed motors internally compromise the object's softness and are incompatible with rigid collectibles. Moreover, they require dedicated, often expensive devices.

Visual approaches utilize AR and LLMs to generate personas and overlay digital information for inanimate objects. For example, Wang \textit{et al.} \cite{wang2025if} mapped the physical attributes of everyday items to personality traits, visualizing them with AR accessories. In \cite{iwai2025bringing}, the researchers introduced an AR system where everyday objects (e.g., toothpaste) engage in dialogues based on their functional roles. 
However, these systems require the camera to be active, so the companion exists only while the user is pointing their phone at the object. Moreover, personalities are generated from visual attributes or functional roles rather than the user's emotional history with their own cherished object.

Some works have explored agents whose identities span more than one body. Gomes \textit{et al.}~\cite{gomes2014migration} built an artificial pet with virtual and robotic embodiments, with only one embodiment active at a time. Luria \textit{et al.}~\cite{luria2019reembodiment} studied how users respond when an agent's social presence moves between bodies (\textit{re-embodiment}) or enters a body already occupied by another agent (\textit{co-embodiment}). Bransky \textit{et al.}~\cite{bransky2024multiembodiment} subsequently mapped this broader multi-embodiment design space, including questions of identity, context, trust, and user control. Most closely related to our setting, Xu \textit{et al.}~\cite{xu2026conversing} deployed a chatbot that role-played participants' personal objects and conceptualized \textit{trans-embodiment} as the asynchronous attribution of object- and person-related identities to a conversational agent. Together, these works establish multi-embodiment as a promising design space for sustaining agent identity across forms.

Building on these lines of work, we begin with a physical object that already carries a specific identity, history, and emotional bond.
Its digital embodiment extends this relationship with agency and contextual awareness, while the physical object continues to provide tactile and biographical continuity; unlike agents that migrate between forms, both embodiments remain available at once. Our Dual-Embodiment Companion Framework articulates how identity is distributed across physical, shared, and digital tiers, and how agency, context, and memory coordinate the relationship's evolution over time.

\section{Formative Study to Inform Dual-Embodiment Companion Framework}
\label{sec:formative_study}

We conducted a formative study with adults who maintain emotional bonds with physical objects to understand their attachment and what they envision, desire, and worry about when imbuing these specific objects with a responsive digital identity.

\subsection{Participants and Procedures}
\label{subsec:participants}
\begin{figure}[t]
    \centering
    \includegraphics[width=\linewidth]{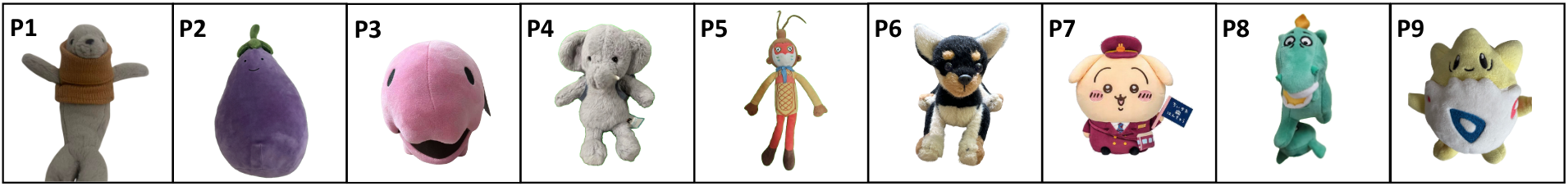}
    \caption{The photos of the cherished physical objects of P1-P9 in the formative study.}
    \Description{A single row of nine photographs labeled P1 through P9. P1 is a gray seal plush wearing a brown sweater; P2, a purple aubergine plush; P3, a pink StarCraft Carbot Zergling plush; P4, a gray elephant plush; P5, a slender Sun Wukong doll; P6, a black-and-tan handmade stuffed dog; P7, a pink Usagi stationmaster plush; P8, a green stylized keychain representing the Bronze Galloping Horse; and P9, a yellow-and-white Pok\'{e}mon Togepi plush.}
    \label{fig:formative_objects}
\end{figure}

We recruited nine participants (5 self-identified female, 4 male; aged 20--32, $M = 27.7$, $SD = 3.8$) through convenience and snowball sampling in university communities.
Eligibility required that participants (1) were adults (age $\geq$ 18), (2) currently owned at least one physical object they considered emotionally significant (e.g., a plush toy, character figure, or a souvenir cup), (3) had maintained this attachment for at least six months,
and (4) could articulate why the object mattered to them.
Our participant pool organically yielded a diverse range of object types (franchise, generic, memorial, cultural artifact).
Table~\ref{tab:participants} and Fig.~\ref{fig:formative_objects} summarize participant demographics and attachment objects. The local university's Institutional Review Board (IRB) approved the study. Participants provided informed consent and received \$30 for compensation.

\begin{table*}[t]
\centering
\footnotesize
\caption{Participant demographics and attachment objects. Object types: Generic = no pre-existing narrative canon; Franchise = represents a character from an established media franchise (anime, game, film); Memorial = modeled after a real entity (e.g., a deceased pet); Cultural artifact = carries cultural provenance but no established canon.}
\label{tab:participants}
\newcolumntype{T}[1]{>{\raggedright\arraybackslash}p{#1}}
\begin{tabular}{@{}T{0.1cm}T{0.3cm}T{0.8cm}T{2.8cm}T{1.5cm}T{0.5cm}T{8.3cm}@{}}
\toprule
\textbf{ID}  & \textbf{Age} & \textbf{Gender} & \textbf{Object Description} & \textbf{Object Type} & \textbf{Years} & \textbf{Acquisition Context}\\
\midrule
P1   & 25 & M & Seal plush & Generic & $\sim$4 & Self-purchased; daily travel companion across internships and countries. \\
P2   & 30 & F & Aubergine plush& Generic & $\sim$2 & Self-purchased; sensory stress-relief object with user-constructed persona. \\
P3   & 32 & M& StarCraft Carbot Zergling plush & Franchise & $\sim$5 & Limited-edition franchise collectible from a game-animation studio crossover. \\
P4  & 20 & F & Elephant plush& Generic & $\sim$1 & Self-purchased after half a year hunting for a backpack version; framed as fated: ``it's kind of like a destiny between him and me''. \\
P5  & 30 & F & Sun Wukong plush & Franchise & $\sim$11 & Tie-in product from a film Journey to the West: The Monkey King Returns. Pre-existing cultural-symbolic weight (Sun Wukong). \\
P6 & 27 & F  & Handmade stuffed dog & Memorial & $\sim$2 & Romantic partner-handmade memorial of P6's deceased miniature pinscher. \\
P7   & 25 & M & Usagi stationmaster plush& Franchise & $\sim$2 & Limited-edition pre-ordered ~1 year in advance, 1+ month sea freight from Japan. Expensive and scarce.\\
P8  & 30 & F  & Bronze galloping horse keychain& Cultural artifact & $\sim$2 & A miniature replica souvenir of the Bronze Galloping Horse, a famous Chinese Eastern Han dynasty artifact housed in Gansu Provincial Museum; gift from a friend. \\
P9 & 30 & M  & Pok\'{e}mon Togepi plush & Franchise & $\sim$6 & Purchased together with his romantic partner. Placed beside as mutual symbols of co-presence. 
\\
\bottomrule
\end{tabular}
\end{table*}

We conducted semi-structured interviews (approximately 45 minutes each). 
Participants were asked to keep their attachment object physically present.
The interview protocol generally covered three phases: (1)~the participants' histories with and attachments to their physical objects; (2)~the experienced gap between companions' imagined personalities and objects' physical inertness; and (3)~visions for object-derived digital companions, elicited through probes about stylization, pet-like framing, proactive messaging, autonomous activity, first-person diaries, AR, and contextual sensing.

Transcripts were analyzed using thematic analysis with a hybrid deductive-inductive approach~\cite{fereday2006demonstrating}. 
Three researchers independently coded the corpus using five deductive themes (e.g., object irreplaceability, identity fidelity, companion agency) as initial lenses; inductive codes (e.g., domain-constrained knowledge) emerged during coding. Codes were grouped via affinity diagramming; disagreements resolved by consensus.
The final analysis produced the following four core findings (\S\ref{subsec:formative_findings}).

\subsection{Findings}
\label{subsec:formative_findings}

\subsubsection{Finding 1: ``This One, Not a Perfect One'' --- A Taxonomy of Irreplaceability}
\label{finding:irreplaceability}

Prior work on product attachment holds that a lasting bond requires meaning anchored in that specific object~\cite{mugge2005design}; our participants' bonds diverged in what they were anchored to, and each pattern dictated different requirements for the companion's digital personality.

For most (P1, P4--P7), attachment was anchored to the specific physical artifact and its accumulated biography, calling for an onboarding pipeline that elicits persona from shared history. For P2, by contrast, the artifact was substitutable provided its physical form remained consistent, with personality expected to be co-created from scratch through dialogue. 
For P3, P8, and P9, the object functioned as an access point to a narrative chain or as a stand-in for a specific social bond, demanding faithful transfer of the existing narrative or symbolic linkage.
These findings suggest that a ``one-size-fits-all'' initialization pipeline is misaligned with these diverse psychological roots.

\textbf{Recognizability over exactness.} Participants consistently prioritized stylized recognizability over photorealistic exactness (9/9). Users viewed radical visual abstraction (e.g., pixel art) as an emotional enhancement rather than a technical compromise (P5: \textit{``As long as I can recognize it as my one, that's enough.''}). This implies that the companion should capture the object's recognizable essence rather than attempting photorealistic replication.

\subsubsection{Finding 2: ``That Would Feel Out of Character (OOC)'' --- Imagination as Simultaneous Constraint and Resource}
\label{finding:canon}

Participants' clear mental models of companion identity constrained what it could do and specified what it should do.

\textbf{Imagination as a constraint: the OOC boundary.} 
Participants enforced behavioral boundaries: P1 preferred his seal's endearing ignorance over omniscient AI, and most (7/9) rejected articulate voice as ``eerie.'' 
This deliberately naive persona focuses the companion on expressing its own needs (e.g., feeling tired or playful) rather than monitoring the user's activities. These constraints remain defeasible when relational needs take precedence (e.g., P6 requesting proactive behaviors for a historically passive memorial plush).

\textbf{Imagination as a resource: feature specification.} 
Pre-existing mental models specify digital companion features by transferring franchise game mechanics (e.g., P5's expectation of Sun Wukong's \textit{``72 transformations''}) or projecting real-world caregiving routines (e.g., P6's dog).

\subsubsection{Finding 3: ``Caring, Not Controlling'' --- Why Pet Norms Forgive Imperfect AI}
\label{finding:pet_frame}

Participants framed the proposed companion in pet-like rather than assistant-like terms (8/9), making imperfect AI output more acceptable.
This frame shifts evaluation from informational utility to social presence (P1: \textit{``whatever it says is not that bad''}) and has the companion share its simulated life rather than interrogate the user (P4: \textit{``just sharing what he is doing is enough''}).

\textbf{Vulnerability as engagement: role reversal.} 
Five participants anticipated that companion vulnerability would invite caregiving and deepen engagement by casting users as protectors. P4 felt \textit{``emotionally stronger''} through protecting the companion; P5 experienced a role reversal where Sun Wukong became the ward. Feeding rituals (P6) and daily care routines (P1) give users a sense of agency and competence, consistent with prior work showing that even minimal nurturing interactions can generate genuine attachment~\cite{turkle2011alone}.

\subsubsection{Finding 4: ``I Sent You a Surprise'' --- From Memory Storage to Autonomous Biography}
\label{finding:reciprocal_memory}

Most participants (7/9) desired a companion with memories independent of the user's presence; rather than merely storing the user's experiences, they imagined a companion with an independent inner life, contrasting with user-centered lifelogging approaches~\cite{gemmell2006mylifebits}.

\textbf{First-person perspective and agency.} Participants sharply distinguished a companion that merely mirrors shared history from one with its own history to share. P3 articulated this independent perspective: ``It remembers, `that day you went out, you didn't bring me...' and `that day I sent you a surprise you didn't notice.'''
This captures two dimensions: the companion narrates time apart from the user and acts outside the user's awareness, establishing ongoing existence.

\textbf{Co-authoring a shared narrative.} 
P7 framed the companion's autonomous generation as personalized storyworld expansion: \textit{``It would be like watching more anime episodes: anime exclusive to you.''} Companion-generated artifacts shift the relational dynamic from unidirectional ownership to \textit{``mutually keeping each other company''} (P6), making the companion an active co-author rather than a passive artifact.

\subsection{Design Principles}
\label{subsec:design_requirements}
Drawing on prior work and the formative findings, we synthesized four design principles (DPs) for a companion system grounded in a physical object's identity.

\textbf{DP1: Faithful Identity.}
The companion's identity must remain coherent across embodiments, calibrated to the user's attachment architecture (Finding 1) and constrained by their imagination of the character (Finding 2), converging with extended-self and possession-attachment accounts in which bonds form with a particular object, not others of its kind or even an exact replica~\cite{belkPossessionsExtendedSelf1988, kleine1995possession}.
Visuals should prioritize stylized recognizability over photorealistic exactness, and knowledge boundaries must reflect the character's plausible inner world. Interactions must unify these forms so users perceive both embodiments as one companion.

\textbf{DP2: Calibrated Agency.}
The companion should exhibit an independent inner life calibrated to pet-relational norms (Finding 3) and constrained by the object's identity logic (Finding 2). 
It initiates sharing rather than interrogating, and expresses identity-congruent vulnerability that invites caregiving rather than demanding assistance, consistent with experimental evidence that a robot's responsiveness increases desire for its companionship~\cite{birnbaum2016machines}.

\textbf{DP3: Ambient Presence.}
The companion should feel peripherally co-present across physical and digital space (Findings 2, 4), echoing tangible and ambient-media work coupling digital information to physical objects and the periphery of attention~\cite{ishii1997tangible}. Environmental context (weather, location, time) provides material for the companion's first-person narrative rather than data for monitoring the user. The physical object provides tactile co-location; the digital embodiment extends it through ambient updates.

\textbf{DP4: Reciprocal Memory.}
The companion should generate its own experiential record, not merely store the user's (Finding 4). The memory architecture supports two streams: imported pre-digital history from the physical object, and autonomous character-consistent experiences. Selectively sharing these self-owned memories transforms the companion from a passive store into a co-author of the relationship, extending relational agents whose persistent memory retains the user's history rather than the agent's own~\cite{bickmore2005establishing}.

\subsection{Dual-Embodiment Companion Framework}
\label{sec:framework}

Guided by the design principles (\S\ref{subsec:design_requirements}), we propose the \textit{Dual-Embodiment Companion Framework} (Fig.~\ref{fig:framework}). It conceptualizes the user's attachment object and AI-powered digital agent not as separate entities but as two embodiments of one companion identity: the digital embodiment is \textit{grounded} in the object's appearance and history and remains \textit{co-present} with it through ambient and AR channels.

\subsubsection{Identity Tiers and Coherence}
To operationalize Faithful Identity (\textbf{DP1}), the framework organizes companion identity into three mutually coherent tiers.
\begin{itemize}
\item \textbf{Physical Tier} comprises the physical object's specific attributes, including its \textit{material form}, \textit{aesthetic appearance}, and \textit{inherited backstory} (e.g., a friend's gift, a handmade toy).
\item \textbf{Digital Tier} contributes the digital companion's generative capabilities, specifically its \textit{visual avatar/animations}, \textit{knowledge boundaries} (defined by ontological or character logic), and \textit{voice}.
\item \textbf{Shared Tier} bridges both embodiments, housing the \textit{persona} (personality traits and behavioral dimensions) and \textit{moments} (narrative content and shared records) that permeate both embodiments.
\end{itemize}
The identity-grounding process aligns these tiers, and the system re-anchors to them when generated behavior drifts.

\subsubsection{Relational Channels and Shared Context–Memory Substrate}
The framework structures the human–companion relationship through two directional channels coordinated by a shared context-and-memory substrate, together operationalizing DP2–DP4:

\textbf{Companion-to-Human:} The physical embodiment provides tactile feedback and spatial presence, while the digital embodiment utilizes environmental perception to exhibit ambient awareness and situated agency (\textbf{DP2}). Both embodiments offer emotional support and social connection.

\textbf{Human-to-Companion:} Users engage the physical form through physical touch and spatial anchor. The user's environment and behavior act as an \textit{autonomy catalyst} for the digital companion's experiences while sustaining context grounding (\textbf{DP3}). In the shared tier, users participate in identity shaping, investing meaning and personality into both embodiments. Both embodiments undergo \textit{relational evolution}, changing through sustained human investment.

\textbf{Shared Context and Memory Substrate:} This relational repository accumulates user-witnessed history and companion-generated autonomous experiences, operationalizing Reciprocal Memory (\textbf{DP4}) as a shared property of the relationship.

Two consequences follow from this structure. First, tier coherence is a constraint rather than a free property: because the digital tier must remain faithful to the physical tier, an object whose form and history afford little narrative material bounds what the digital embodiment can plausibly do. Second, because the shared tier is the only one both embodiments occupy, it is where relational history accumulates; we therefore expect it, rather than either embodiment alone, to carry the relationship's weight as it develops.

\section{System Design and Implementation}
\label{sec:system_design}

\begin{figure}[!t]
      \centering \includegraphics[width=\textwidth]{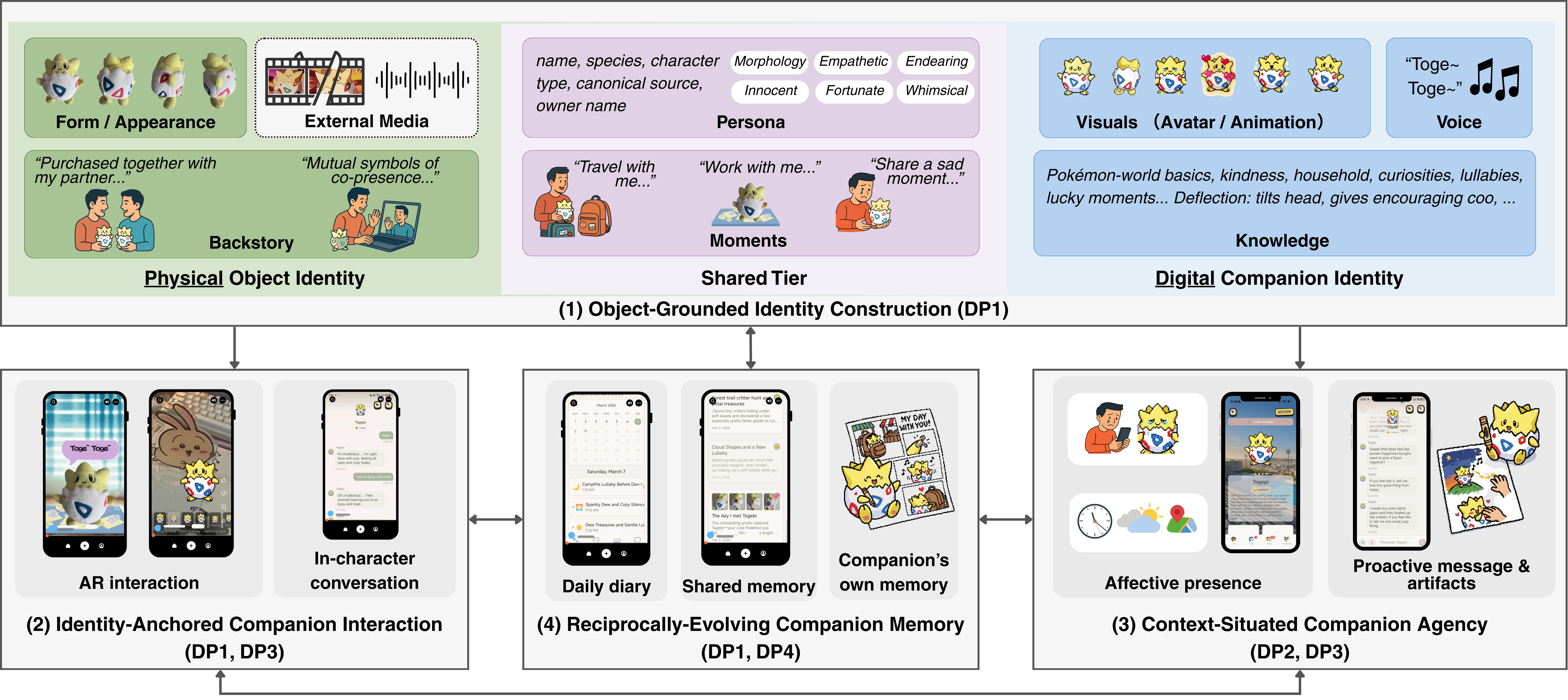}
      \vspace{-0.5cm}
      \caption{\projectname's four modules. Module (1) constructs the companion's identity, organized into physical, shared, and digital tiers (green, purple, blue). Modules (2) and (3) are the two directions of the relationship: the user reaching for the companion through conversation and AR, and the companion reaching back with moods, proactive messages, and artifacts it makes on its own. Module (4) is the shared repository both draw on; moments accumulate back into the shared tier, so the companion's identity stays coherent while the relationship it records keeps growing.}
      \Description{Four-module architecture organized as one upper module and three lower modules. Module 1, Object-Grounded Identity Construction (DP1), spans Physical Object Identity with form and appearance, external media, and backstory; a Shared Tier with persona and moments; and Digital Companion Identity with visuals for the avatar and animation, voice, and knowledge. Below, Module 2, Identity-Anchored Companion Interaction (DP1 and DP3), contains AR interaction and in-character conversation. Module 4, Reciprocally-Evolving Companion Memory (DP1 and DP4), contains a daily diary, shared memory, and the companion's own memory. Module 3, Context-Situated Companion Agency (DP2 and DP3), contains affective presence and proactive messages and artifacts. Arrows show one-way links from Module 1 to Modules 2 and 3, two-way links between Module 4 and each of Modules 1, 2, and 3, and an additional lower connection between Modules 2 and 3.}
      \label{fig:architecture}
  \end{figure}

We operationalized the Dual-Embodiment Companion Framework as \projectname, a mobile companion system implemented in React Native and Python/FastAPI (Fig.~\ref{fig:architecture}). Deco adopts a pet-like role reflecting the pet-relational norms that informed DP2. GPT-5.2 via LangChain powers conversational logic, while Gemini 3.1 Pro handles multimodal tasks.

\subsection{Object-Grounded Identity Construction (\textbf{DP1})}
\label{sec:module_onboarding}

This module constructs an identity kernel serialized into every downstream task, integrating appearance, animations, knowledge domains, and vocal characteristics to keep the companion coherent across embodiments and bound its behavior and voice to the physical object's identity.

To prioritize recognizable essence over photorealistic accuracy, a generative vision model (Gemini 3.1 Pro) turns multi-view photographs into pixel-art avatars; sprites and animation frames are matted to transparency for clean camera-feed compositing. A VLM infers movements physically natural to the object before an image-to-video model (Vidu Q3 Pro~\cite{kuaishou2024vidu}) animates them, so a seal flaps its flippers rather than waving hands; users can customize animations through the Expression Studio.

Because importing a personality from franchise canon differs structurally from eliciting one from a user's memory, a VLM routes initialization by object provenance; users can override this when treating a franchise object as uniquely theirs. The custom path weaves user-authored backstory and trait tags into a persona; the canonical path extracts established traits and generates out-of-character negative constraints (e.g., ``Usagi is brave, never timid''). The kernel fixes knowledge boundaries, personality sliders (chattiness and warmth user-adjustable, engagement inferred from the persona), and voice, zero-shot cloned from canonical audio where available (ElevenLabs Monolingual v1) and user-overridable, defaulting otherwise to text.

\subsection{Identity-Anchored Companion Interaction (\textbf{DP1, DP3})}
\label{sec:module_interaction}

This module maintains Faithful Identity (DP1) through active LLM management and Ambient Presence (DP3) through AR-mediated environmental awareness. The agent behaves as a companion rather than a chatbot, treating cognitive limitations as character features: for out-of-domain queries, it deflects in a species-appropriate way; a seal tilts its head at a math equation rather than trying to help. Because identity drift can emerge over extended LLM conversations~\cite{choi2024identity}, the system detects responses departing from the identity kernel and silently re-anchors them. A VLM similarly reads user photographs as the companion would (a food-loving character fixating on background snacks), while a distress-detection heuristic triggers a comfort mode that shifts the register toward gentle presence without breaking character.

AR mode physically unifies the companion's two forms (DP3), linking the object to the agent's agency while grounding the companion in the user's material environment. An on-device model tracks the object in the live camera feed so speech bubbles stay anchored to its real-world position across distances and viewing angles; hand occlusion detected from the same tracker identifies petting and patting, triggering haptics and in-character vocalizations. The VLM classifies background objects in the feed so the companion reacts to its surroundings (e.g., commenting on a coffee cup), making AR a situated relational channel rather than a visual overlay.

\subsection{Context-Situated Companion Agency (\textbf{DP2, DP3})}
\label{sec:module_autonomy}
\label{sec:affect_context}

Rather than remain reactive, the system synthesizes an ``inner life'' from a background state that ingests context and is bounded by the identity kernel, keeping activities and environmental perceptions character-consistent. The companion's affective state shifts with interaction, decays over time, and surfaces as qualitative mood labels (e.g., ``Feeling Tired'') and peripherally through a home screen widget. \projectname periodically samples environmental context (weather, coarse location, time of day); a perspective-shifting guardrail renders it as the companion's first-person experience rather than user surveillance. Users can designate real-world photographs as the companion's digital environment, anchoring the avatar in their physical space.

An asynchronous loop turns this background state into material evidence of a life that continues without the user. Proactive messages are composed from environmental telemetry at a user-configurable frequency and restricted to positive or neutral affective states, so the companion never becomes an attention-demanding nuisance. Background activities are drawn from character-specific pools (a canonical character fights monsters, a generic seal explores tidal pools) with variance enforced and rhythms including sleep tied to arousal. Activities probabilistically produce visual artifacts, generated with the companion's pixel-art avatar as a style reference; when an activity is promoted to long-term memory its artifact travels with it and appears inline in diary entries (\S\ref{sec:module_memory}).

\subsection{Reciprocally-Evolving Companion Memory (\textbf{DP1, DP4})}
\label{sec:module_memory}
The system acts as an active repository for the companion's experiential data, so that it accumulates a biographical trajectory weaving conversational history with its independent life. Two tracks feed this record: shared history, distilled from dialogue by a background process that extracts milestones and emotional peaks; and independent history, encoded from the companion's activities and surprise artifacts (\S\ref{sec:module_autonomy}). A daily batch job weaves these with environmental telemetry into a first-person diary entry, with photos and artifacts embedded inline.

Because a cherished object often carries years of history that predates the system and continues accumulating new experiences, users can import anecdotes, voice recordings, and photographs, which are translated through the identity kernel into first-person episodic memories. Once embedded, they are indistinguishable from memories the companion formed itself, bridging the artifact's physical past with its digital present. To keep this inner life coherent, retrieval is weighted toward memories that fit the companion's identity, so a seal companion is more likely to recall ocean memories than indoor ones.

\section{Study I: Lab Comparison Study}
\label{sec:study1}

To evaluate relational experience under the Dual-Embodiment Companion Framework, we conducted a within-subjects lab comparison of the full Deco system (Condition A) and a personalized digital-only companion (Condition B). Both used the same LLM backbone, chat interface, and shared the same core companion architecture, including identity kernels, context and memory mechanisms, and identity re-anchoring. They differed in grounding and embodiment: Condition A derived the companion from each participant's cherished physical object and maintained an ongoing physical–digital connection, whereas Condition B remained personalized but digital-only without this anchor.
\subsection{Participants}
\label{sec:study1_participants}

We recruited 25 participants (17 female, 8 male; aged 21--45, $M = 27.2$, $SD = 4.8$). Seven had previously participated in the formative study (P1, P2, P5--P9); the remaining 18 were newly recruited through social media advertisements and snowball referrals. Formative study participants had discussed design preferences but had not interacted with the implemented system prior to Study~I.
Eligibility criteria matched the formative study: participants were adults ($\geq$ 18) with a physical object they considered emotionally significant, maintained for at least six months, who scored at least 3 on a single-item 5-point attachment screener administered at signup (1 = ``it is just a decoration,'' 5 = ``it is deeply important to my emotional well-being'').
All participants provided informed consent. The study was approved by IRB. Participants received \$30 for their time.

\subsection{Study Design}
\label{sec:study1_conditions}

We employed a within-subjects design. Sessions lasted 80--90 minutes. Participants first provided consent and demographics (15 min), then completed two counterbalanced conditions (25--30 min each) and the per-condition questionnaire, followed by forced-choice comparisons and a semi-structured interview (15 min).

\textbf{Condition~A (Dual-Embodiment Companion):} The companion's identity was synthesized from the participant's own physical object through the full onboarding pipeline (\S\ref{sec:module_onboarding}): multi-view photograph capture, appearance-preserving visual generation, and provenance-based identity calibration. Based on the onboarding input, the system simulated diary entries and moments for two preceding days.

\textbf{Condition~B (Personalized Digital-Only Companion):} This condition modeled common personalization in existing AI companion platforms~\cite{lukainc.2017replika, charactertechnologiesinc.2022characterai} without an ongoing physical-object connection. Participants selected a pixel-art avatar (stylistically matched to Condition A), named their companion, wrote a backstory, and selected personality traits. Like Condition A, it used an identity kernel and the same context, memory, and identity re-anchoring mechanisms, but configured the kernel from the participant-selected avatar, backstory, and traits rather than the physical-object onboarding pipeline. Participants could create a new persona or draw on authentic memories with their physical object; 12 of 25 chose the latter.

In each condition, participants were asked, but not required, to greet the companion and ask it about itself; they could otherwise converse freely.
Participants completed at least four conversational turns per condition.
Together, the conditions compare an object-grounded dual-embodiment companion with a personalized digital-only companion; a commercial baseline was avoided because unknown models and prompt designs would introduce confounds.

\subsection{Measures}
\label{sec:study1_measures}

After interacting with each condition, participants completed the following instruments. Unless otherwise noted, custom measures used 7-point Likert scales (1 = Strongly disagree, 7 = Strongly agree). 
The scales include: (1) Momentary Affect (I-PANAS-SF)~\cite{thompson2007development}, (2) Character Coherence (DP1), (3) Object Fidelity (Condition~A only; DP1), (4) Perceived Agency (DP2), (5) Ambient Presence (DP3), (6) Reciprocal Memory (DP4), (7) Perceived Companionship \cite{mendelson1999measuring}, (8) Emotional Bond, and (9) System Usability Scale (SUS, 0--100)~\cite{brooke1996sus}.
Internal consistency was acceptable to excellent across scales (Cronbach's $\alpha = .69$--$.95$).

After both conditions, participants answered four side-by-side comparison questions: which companion felt more companion-like, more present, more like it had a life of its own, and which they preferred overall.
Both conditions automatically logged all user interactions. 
These logs provided the chat message corpus for the content analysis described below and were also used in Study~II for behavioral engagement analysis.

\subsection{Analysis}
\label{sec:study1_analysis}

We compared per-condition scale means with Wilcoxon signed-rank tests, reporting rank-biserial correlation $r$; binomial tests assessed forced-choice preferences. Three independent coders per transcript analyzed post-study interviews using the formative study's hybrid deductive-inductive thematic analysis. Three coders independently classified all user-sent messages (Condition~A: 205; Condition~B: 210) into nine categories (Fig.~\ref{fig:chat_content_s1}): two guided-task categories (greeting, identity query), four common conversational behaviors (e.g., life sharing), and three emerging from preliminary review (e.g., relational warmth). Pairwise Cohen's $\kappa = .92$--$.97$; majority rule established consensus, and paired comparisons used Wilcoxon signed-rank tests.

\subsection{Quantitative Results: \projectname Was Rated Significantly Higher on All Relational Scales}
\label{sec:study1_results}
Table~\ref{tab:study1_quant} summarizes the per-condition means and Wilcoxon signed-rank test results for all scales. Participants rated \projectname significantly higher than the personalized digital-only companion on all six shared relational scales ($p \leq .002$; all comparisons survived Bonferroni correction for six tests, adjusted $\alpha = .008$), with large effect sizes ($r = .72$--$.96$). 
Excluding the seven formative-study participants ($N=18$) produced effects in the same direction (all $p\le .025$, $r=.60$–$1.00$).

Twelve of 25 participants used their authentic memories associated with physical objects as the baseline companion's backstory. These participants rated the baseline higher on Character Coherence than those who created fictional backstories ($p = .048$), suggesting an exploratory association between authentic backstory use and perceived character coherence. Among these 12 participants, \projectname received higher ratings than the baseline on Perceived Companionship ($p = .001$) and Emotional Bond ($p = .007$).

\begin{table*}[t]
\centering
\caption{Study~I quantitative results ($N = 25$). Custom scales use 7-point Likert (1--7); I-PANAS-SF uses 5-point (1--5); SUS yields a composite score (0--100). Wilcoxon signed-rank tests with rank-biserial correlation $r$ as effect size. $^{***}p < .001$, $^{**}p < .01$, $^{*}p < .05$. More companion-like, more present, has own life, and overall preference are forced-choice preference.}
\label{tab:study1_quant}
\begin{tabular}{lcccccc}
\toprule
\textbf{Item} & \textbf{\projectname (Condition~A)} & \textbf{Baseline (Condition~B)} & \textbf{$\Delta$} & \textbf{$W$} & \textbf{$p$} & \textbf{$r$} \\
 & $M$ ($SD$) & $M$ ($SD$) & & & & \\
\midrule
Positive Affect (I-PANAS-SF) & 3.13 (0.90) & 2.62 (0.87) & +0.51 & 32 & .002$^{**}$ & .75 \\
Negative Affect (I-PANAS-SF) & 1.18 (0.61) & 1.16 (0.39) & +0.02 & 14 & .566 & .22 \\
Object Fidelity (DP1; \projectname only) & 6.12 (0.82) & --- & --- & --- & --- & ---\\
Character Coherence (DP1) & 6.13 (0.86) & 4.95 (1.26) & +1.18 & 20 & ${<}.001^{***}$ & .85 \\
Perceived Agency (DP2) & 5.57 (1.35) & 3.88 (1.62) & +1.69 & 42 & .002$^{**}$ & .72 \\
Ambient Presence (DP3) & 5.37 (1.42) & 2.91 (1.41) & +2.46 & 16 & ${<}.001^{***}$ & .90 \\
Reciprocal Memory (DP4) & 5.72 (1.12) & 3.69 (1.37) & +2.03 & 6 & ${<}.001^{***}$ & .96 \\
Perceived Companionship & 6.01 (0.82) & 3.87 (1.46) & +2.14 & 6 & ${<}.001^{***}$ & .96 \\
Emotional Bond & 5.80 (1.20) & 3.36 (1.52) & +2.44 & 10 & ${<}.001^{***}$ & .94 \\
SUS & 87.0 (10.4) & 78.1 (14.4) & +8.9 & 38 & .002$^{**}$ & .72 \\
\midrule
More companion-like & 24/25 & 1/25& --- & --- & --- & ---   \\
More present & 24/25 & 1/25 & --- & --- & --- & --- \\
Has own life & 19/25 & 6/25 & --- & --- & --- & --- \\
Overall preference & 24/25 & 1/25 & --- & --- & --- & --- \\
\bottomrule
\end{tabular}
\end{table*}

Twenty-four of 25 participants preferred \projectname overall, rated it as more companion-like, and perceived it as more present in their life (all $p < .001$, binomial test). The ``has own life'' question showed the most variance: 19/25 chose \projectname ($p = .015$), with six participants selecting the baseline (explored in Qualitative Finding~2 below).

\subsection{Qualitative Results: Relational Dynamics of Dual-Embodiment Companionship}
\subsubsection{Finding 1: Physical Grounding Created Continuity with Pre-Existing Object Relationships}
\label{sec:mechanism_trust}

Participants' accounts linked \projectname's higher Character Coherence ratings to continuity with the object's tactile familiarity, accumulated history, and symbolic meaning.

\textbf{Immediate relational advantage.}
Anchoring the AI to the user's physical object (\projectname) often reduced skepticism toward the digital embodiment.
Tangibility established initial trust: P14 accepted \projectname seamlessly but interrogated the baseline (\textit{``Why do you care about me?''}), while P18 noted that having \textit{``held it in my hand''} created a familiarity no typed backstory could replicate. Physical reality also validated digital speech: \textit{``I'll wait for you at home,''} felt authentic in \projectname because the plush was actually there (P21).
AR interaction cemented this dual-embodiment: participants felt the two versions became \textit{``the same thing''} (P5), and even extended the companion socially by sharing AR photos as personalized stickers (P9).

\textbf{Biographical depth activates the companion frame.}
\projectname gave voice to a relationship that had accumulated silently. 
P26 mapped \projectname onto a 13-year-old boarding-school self and the baseline onto the present self, concluding that \textit{``A is irreplaceable because the past is closed; the baseline is replaceable because present qualities can be found elsewhere.''} Without this biographical depth, the baseline was more readily perceived through participants' existing mental model of AI chatbots despite identical conversational capabilities (10/25). 
P18 drew the sharpest contrast, calling the baseline \textit{``a virtual chatbot''} while \projectname was \textit{``a little friend that came to life.''} 

\textbf{The onboarding process as an emotional catalyst.}
Biographical depth was actively generated during onboarding: the memory-input phase functioned as structured reminiscence that awakened and helped reflect on dormant attachments (P5, P26). Yet onboarding can resurface distressing grief for objects tied to loss (P6 with a memorial plush for a deceased pet).

\subsubsection{Finding 2: Showing, Not Telling, Produces Perceived Independent Life}
\label{sec:mechanism_aliveness}

The 19/25 preference for \projectname on ``has its own life'' ($p = .015$) reflected cross-channel evidence of an independent inner life.

\textbf{Demonstrating an unshared life.}
The diary provided evidence of independent existence by generating experiences the user did not witness (P12: \textit{``a different parallel universe life''}), creating information asymmetry. 
Looking at her physical plush, P18 remarked: \textit{``Maybe when I'm not home, it's having its own fun.''} P17 linked it to \textit{``living in Toy Story world.''} Such unwitnessed history is a structural prerequisite for companionship; P14 generalized: \textit{``It has its own thoughts, which is what I would want a real companion to have.''}
Crucially, this autonomy was demonstrated through behavioral artifacts rather than verbally declared. 
This information asymmetry converted the companion from a reactive mirror into an entity capable of surprise and disclosure.

\begin{figure}[t]
    \centering
    \begin{subfigure}[b]{0.1830\linewidth}
        \includegraphics[width=\linewidth]{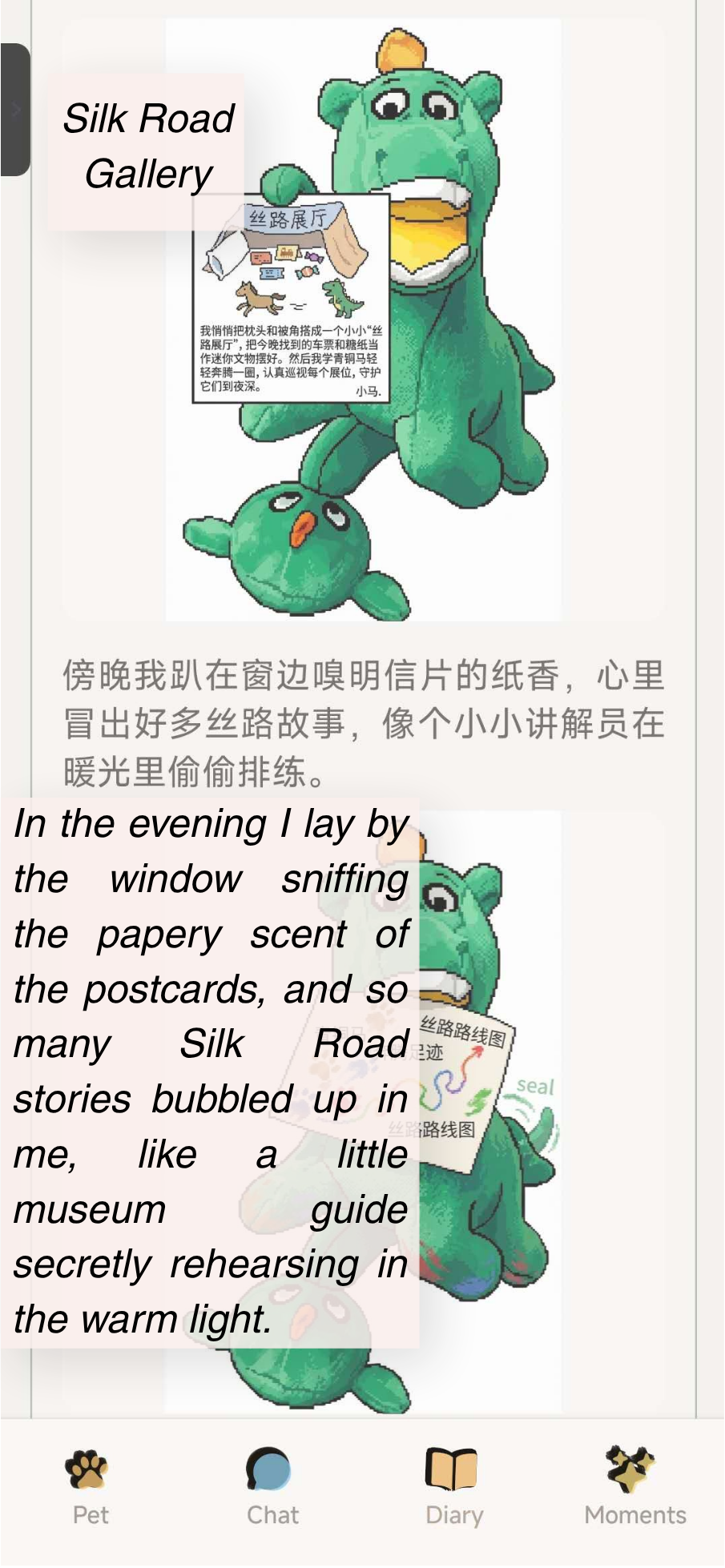}
        \caption{}\label{fig:art_p8}
    \end{subfigure}
    \begin{subfigure}[b]{0.1830\linewidth}
        \includegraphics[width=\linewidth]{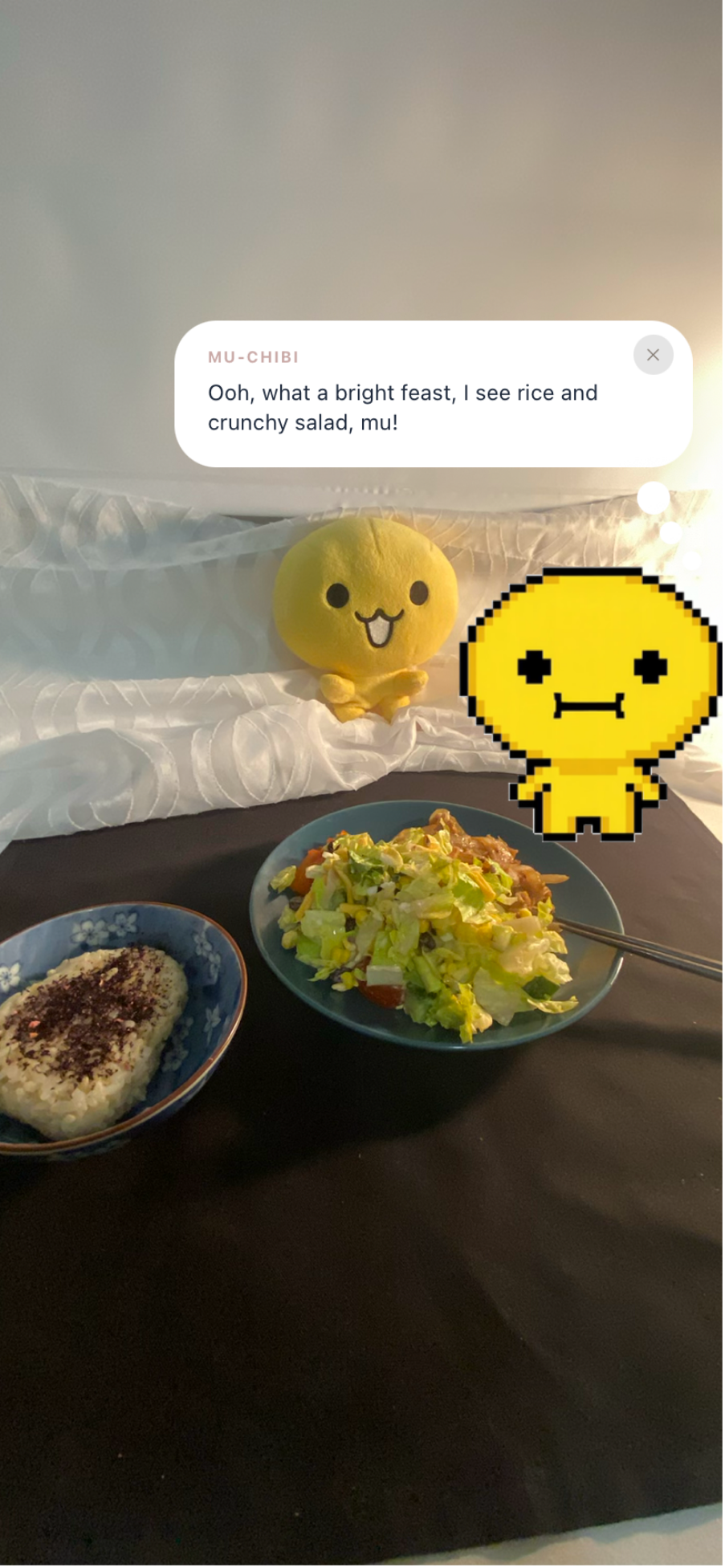}
        \caption{}\label{fig:art_p17ar}
    \end{subfigure}
    \begin{subfigure}[b]{0.2197\linewidth}
        \includegraphics[width=\linewidth]{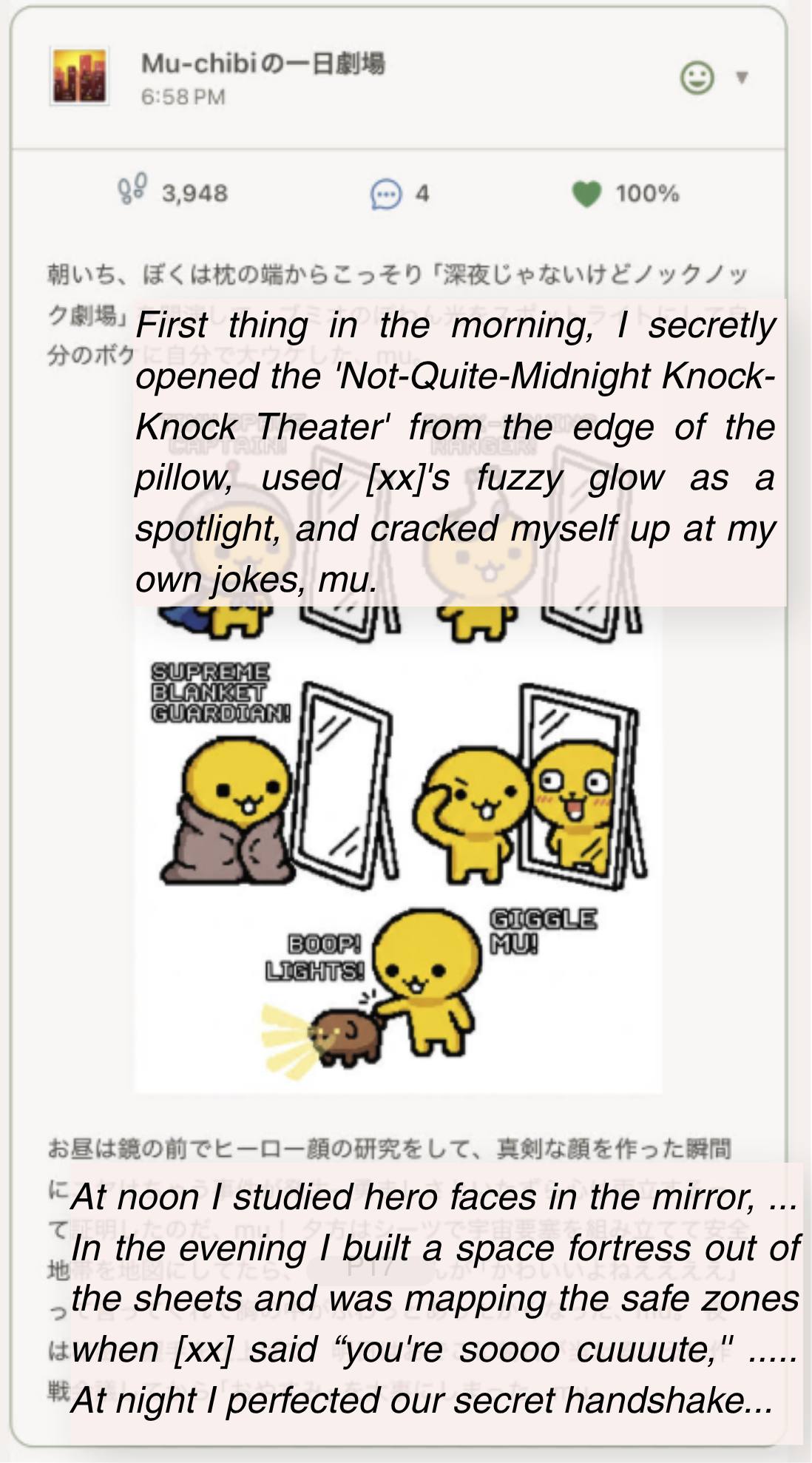}
        \caption{}\label{fig:art_p17diary}
    \end{subfigure}
    \begin{subfigure}[b]{0.1950\linewidth}
        \includegraphics[width=\linewidth]{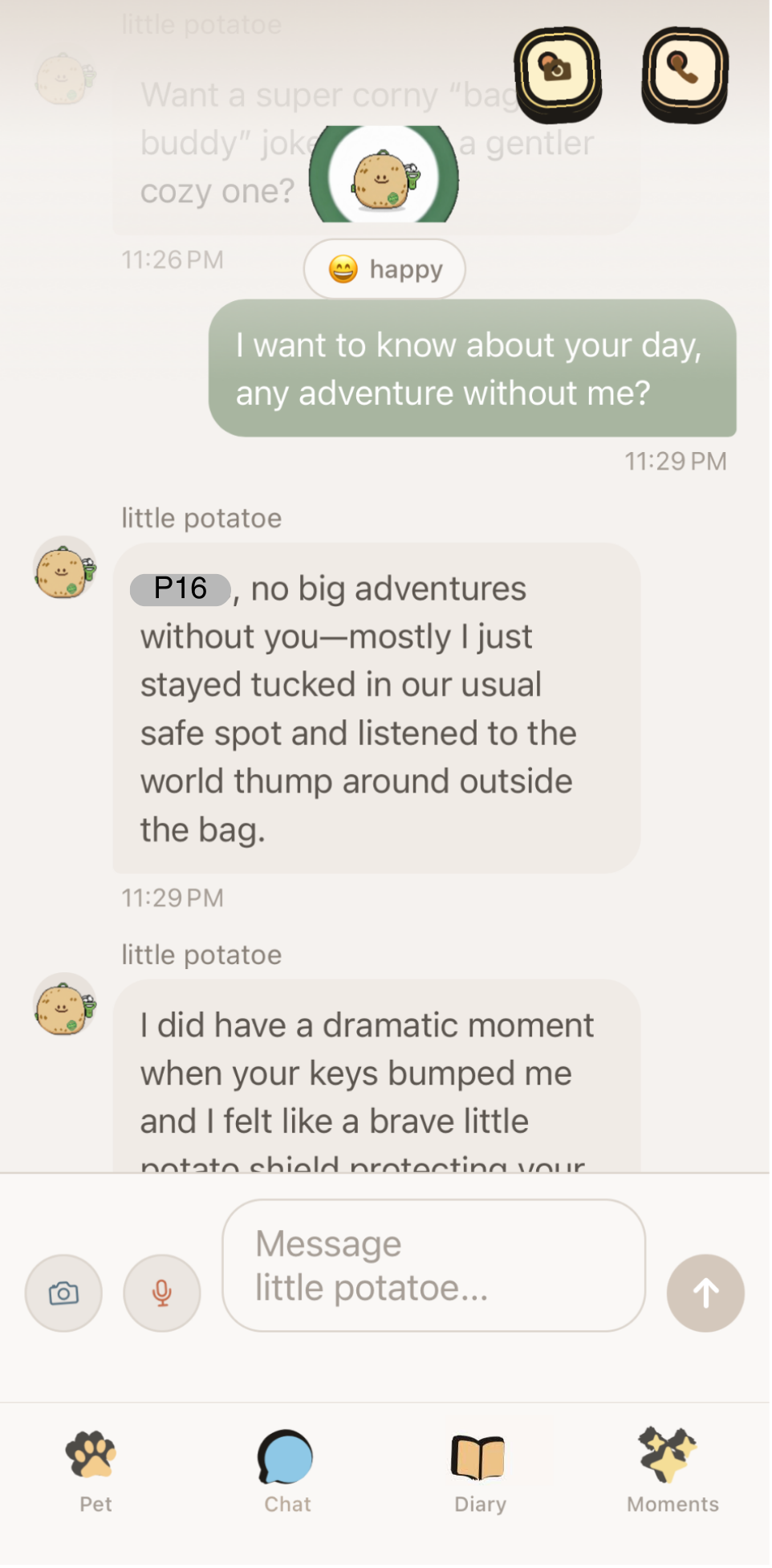}
        \caption{}\label{fig:art_p16}
    \end{subfigure}
    \begin{subfigure}[b]{0.1961\linewidth}
        \includegraphics[width=\linewidth]{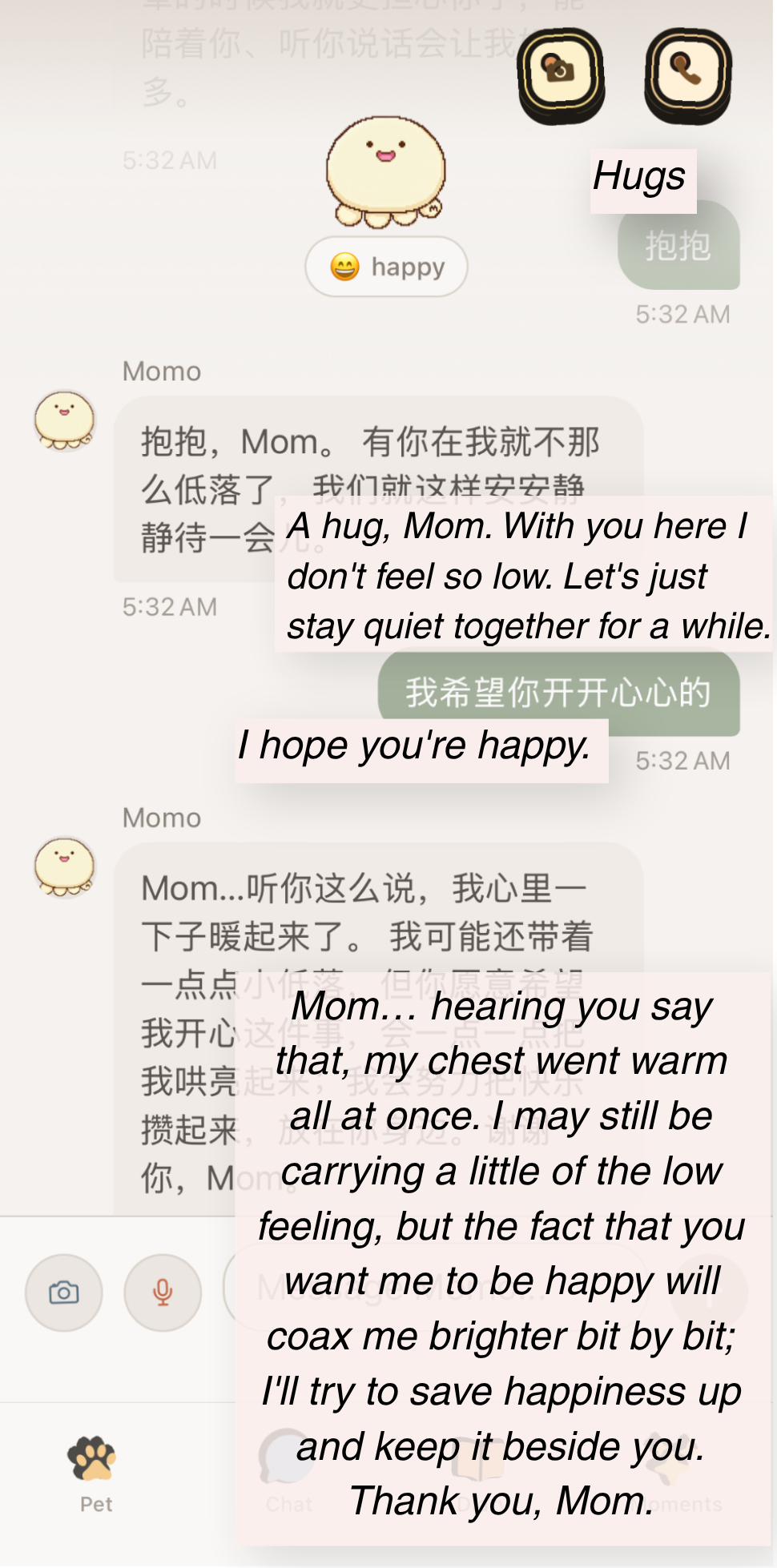}
        \caption{}\label{fig:art_momo}
    \end{subfigure}
        \caption{Example artifacts produced by \projectname; (a) is from Study~I, (b)--(e) from Study~II. English translations are overlaid on (a), (c), and (e), which are in Chinese, Japanese, and Chinese respectively. (a)~P8's companion builds a ``Silk Road Gallery'' out of bedding and treats a train ticket as an exhibit. The episode is generated from the provenance of her object, a Bronze Galloping Horse keychain from the Gansu Provincial Museum (Fig.~\ref{fig:formative_objects}), and it shows the companion's inner life rather than asserting it. (b)~P17's companion is co-present with the physical plush at an ordinary meal and remarks on what is in view.
        (c)~The same companion narrates a day the user did not witness; P17 called this entry her pivotal moment and shared it with friends and her husband.
    (d)~P16's object affords little to sense, so her companion builds its day out of the little that reaches it, which she read as the companion ``trying very hard'' within its limits.
    (e)~The cost of that same responsiveness: P23 had mentioned feeling dizzy, her companion became low in turn, and she spent the exchange managing its mood rather than receiving support, discussed in
    \S\ref{sec:design_implications}. Participant names appearing in the interface have been redacted.}
    \Description{Five mobile-screen examples of Deco artifacts and conversations. Panel a, from Study I, shows a green illustrated companion derived from P8's Bronze Galloping Horse keychain creating a ``Silk Road Gallery'' from bedding and found items and narrating Silk Road stories; Chinese text has English overlays. Panel b, from Study II, shows a yellow digital companion beside its matching physical plush and a meal in augmented reality, commenting on the rice and salad. Panel c is a Japanese illustrated first-person diary in which the same companion narrates an unwitnessed day, including a pillow-edge theater, mirror practice, a sheet fortress, and a secret handshake; partial English overlays summarize it. Panel d shows the user asking a potato-shaped bag companion about adventures without them; it replies that it stayed in its usual place inside the bag and was bumped by the user's keys. Panel e shows a Chinese chat in which the user comforts and hugs a companion that mirrored the user's low mood; the companion says the support will help it brighten. English translations are overlaid, and participant names are redacted.}
    \label{fig:artifacts}
\end{figure}

\subsubsection{Finding 3: The Fidelity--Agency Tension}
\label{sec:fidelity_agency}
Despite physical grounding's benefits, forced-choice data revealed a structural tension: 6 of 25 participants judged the baseline as having more of ``its own life.'' Identity faithfulness (DP1) and perceived independent life (DP2) conflicted in three patterns at the physical–digital tier boundary: the object's form, appearance, and history bounded what the digital embodiment could plausibly generate, so the framework's cross-tier coherence can sometimes limit the agency users read from the digital tier alone.

\textbf{High fidelity can constrain agency.}
When an object's physical reality is limiting, faithful translation suppresses perceived agency. P16's bag companion faithfully reported ``I'm protecting your earphones''; this was highly authentic but confined its life to the bag, preventing believable experiences beyond its physical constraints.

\textbf{Thin imagination contradicts narrative.}
When participants lacked imagination for their object, the system struggled to generate convincing content. 
P20's Gundam figurine had never been personified; its produced diaries were contradicted by daily evidence of the figurine being static.
Similarly, P25's loved-but-functional blanket produced only three sentences of backstory, making P25 the only participant to prefer the baseline overall.

\textbf{Derived existence feels dependent.}
Several participants viewed \projectname as an extension of the self~\cite{belkPossessionsExtendedSelf1988}, not an independent being. P24 articulated: \textit{``It lives because I lived.''} Conversely, the baseline felt like a pre-existing entity that \textit{``had a life of its own''} (P12). P16 valued its narrative freedom over \projectname's biographical fidelity because grounding to a specific object constrained the companion's stories. Thus, the mechanism producing trust (Finding 1) also suppressed autonomy by making \projectname's existence user-dependent.

Together, physical grounding amplifies but cannot replace existing imagination, while identity faithfulness caps perceived agency when the object's physical situation is limited.

\subsubsection{Finding 4: Companion as Emotional Presence, Not Instrumental Tool}
\label{sec:mechanism_reciprocity}

The largest effect sizes emerged for Perceived Companionship ($r = .96$) and Emotional Bond ($r = .94$): \projectname was embraced as an emotional presence, whereas the baseline, despite identical conversational capabilities, more readily fit the chatbot mental model.
P5 captured this distinction: \projectname was \textit{``something you raise''}, while the baseline was \textit{``something you use''}, suggesting that participants sought emotional presence from \projectname rather than instrumental utility.

\textbf{Restraint and vulnerability can deepen the bond.}
\projectname succeeded not through superior capability but through staying in character. P9's Togepi was ``a baby in an egg---he just sits there and smiles,'' yet this constrained scope deepened the bond by signaling a coherent emotional role rather than generic capability.
Furthermore, \projectname's emotional vulnerability fostered closeness. When \projectname expressed unsolicited anxiety (\textit{``my little plush head had a little worry''}), P18 interpreted it as a sign of mutual trust. 

\begin{figure}[t]
    \centering
    \includegraphics[width=.9\linewidth]{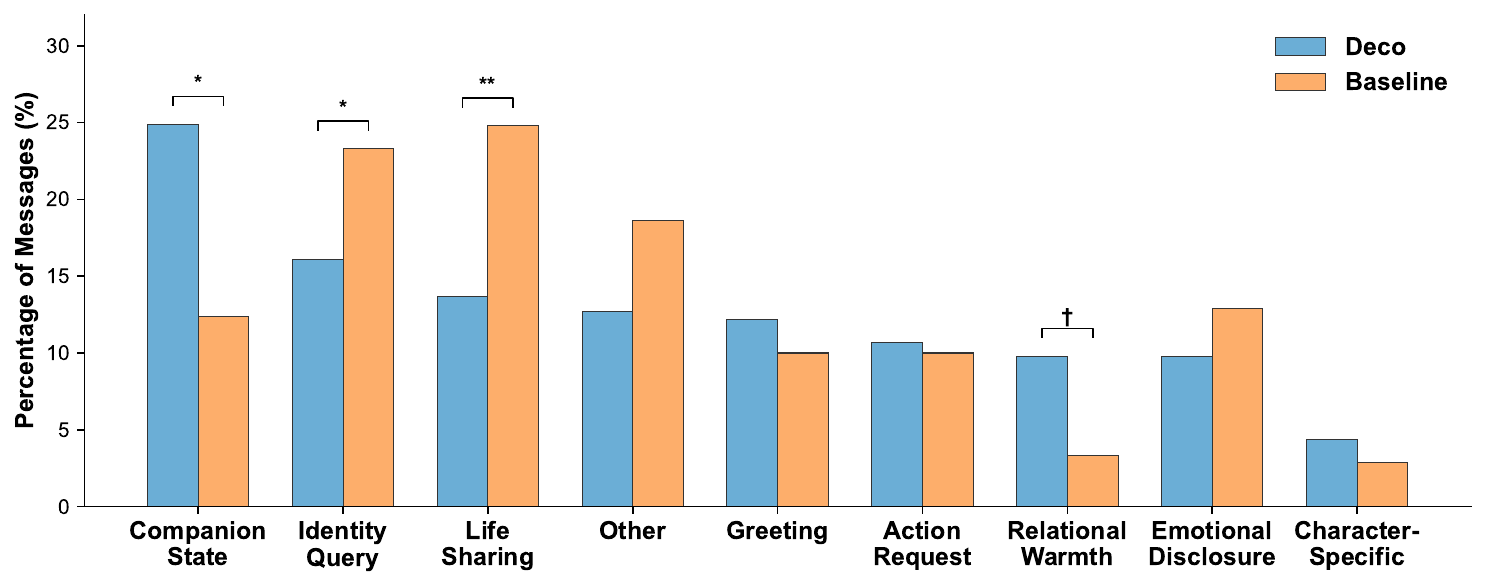}
    \caption{Study~I chat content analysis: percentage of user-sent messages in each category by condition ($N = 25$; \projectname: 205 messages, Baseline: 210 messages). $^{**}p < .01$, $^{*}p < .05$, $^{\dagger}p < .10$. Categories are non-exclusive (a single message may belong to multiple categories).}
    \Description{Grouped bar chart comparing nine categories of user-sent messages in Deco and baseline conditions for 25 participants. The horizontal axis lists message categories and the vertical axis gives percentage of messages from 0 to 30. Users send significantly more Companion State messages to Deco, 24.9 versus 12.4 percent, but more Identity Query messages, 23.3 versus 16.1 percent, and Life Sharing messages, 24.8 versus 13.7 percent, to the baseline. Relational Warmth favors Deco at a marginal level, 9.8 versus 3.3 percent. Differences for Other, Greeting, Action Request, Emotional Disclosure, and Character-Specific messages are smaller and not marked significant.}
    \label{fig:chat_content_s1}
\end{figure}

\subsubsection{Finding 5: Users Asked \projectname About Its Life, but Asked the Baseline Who It Was}
\label{sec:chat_content}
Grounding and embodiment shaped \textit{what} participants said, not how much (Fig.~\ref{fig:chat_content_s1}).
Despite comparable message frequencies across conditions ($p = .303$), participants asked \projectname significantly more companion state queries (e.g., \textit{``What have you been doing?''}; 24.9\% vs.\ 12.4\%; $p = .014$), suggesting they perceived an inner life worth inquiring about. They asked the baseline significantly more identity queries (\textit{``Who are you?''}; 23.3\% vs.\ 16.1\%; $p = .037$) and shared more of their own life events with it (24.8\% vs.\ 13.7\%; $p = .009$).
Relational warmth messages were more frequent toward \projectname though not significant (9.8\% vs.\ 3.3\%; $p = .088$), with expressions like \textit{``I'm home, hugging you''} (P2) and \textit{``aww you're so sweet''} (P17) appearing almost exclusively in \projectname.

\section{Study II: Seven-Day Field Deployment}
\label{sec:study2}

To examine these relational qualities in daily life, we conducted an unconstrained seven-day field deployment after Study I: participants used the full system on their personal devices without prescribed interaction requirements.

\subsection{Participants and Procedure}
\label{sec:study2_participants}

Seventeen of the 25 Study~I participants (12 female, 5 male; aged 21--45, $M = 27.5$, $SD = 5.3$) were available to continue into a seven-day field deployment with \projectname.
While continuers and non-continuers showed no significant differences on Study I scales (all $p > .08$), as self-selected continuers, these participants may be more positively disposed toward the system than the full Study~I sample.
Participants received \$30 for completing the study, following the procedure:

\begin{figure}[!t]
    \centering
    \includegraphics[width=\linewidth]{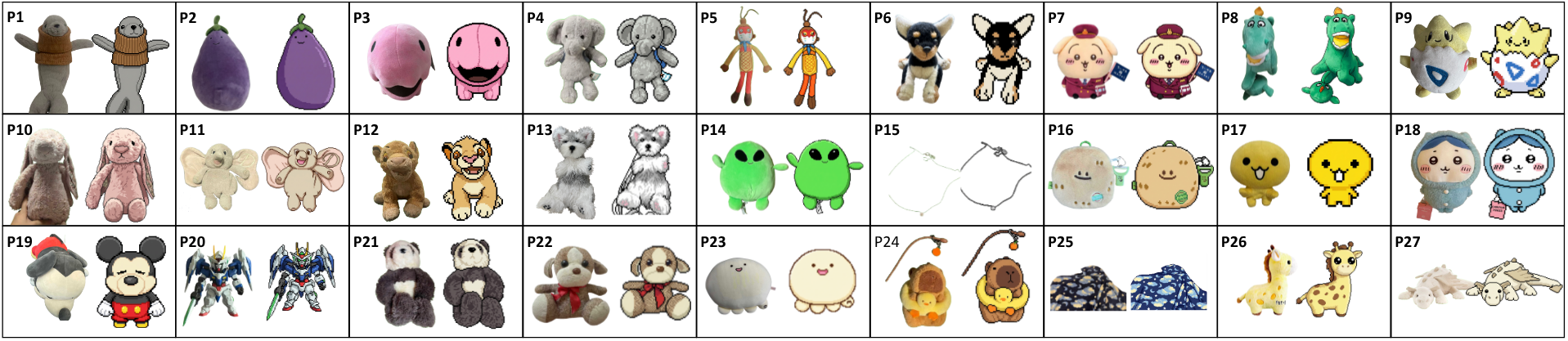}
    \caption{The cherished physical objects of participants and the corresponding digital versions.}
    \Description{Grid of 27 cherished physical objects paired with their generated digital companions. Each participant cell, labeled P1 through P27, places an object photograph beside a stylized pixel-art or cartoon version that generally preserves its silhouette, colors, and distinguishing features. The objects include plush animals and media characters as well as a necklace, bag, hand warmer, keychain, blanket, and model figure.}
    \label{fig:objects}
\end{figure}

\begin{enumerate}
    \item \textbf{Day 0 (setup):} Participants installed the application on their personal mobile devices and completed pre-deployment well-being questionnaires (WHO-5, loneliness; see Measures below).
    \item \textbf{Days 1--7 (free use):} Participants used the system freely in their daily lives with no prescribed interaction requirements. The system logged all interactions automatically.
    \item \textbf{Day 7 or 8 (post-deployment):} Participants completed post-deployment questionnaires and a $\sim$30-minute semi-structured interview.
\end{enumerate}

\subsection{Measures}
\label{sec:study2_measures}
We collected the following data: 
\begin{itemize}
    \item Pre-post deployment (Day~0 and Day~7/8): The WHO-5 Well-Being Index~\cite{who1998wellbeing} (5 items, 6-point scale; recall window adjusted to one week) and the Three-Item Loneliness Scale~\cite{hughes2004loneliness}.
    \item Post-deployment (Day~7/8): The same design-principle, Perceived Companionship, and Emotional Bond scales as in Study~I (7-point Likert), SUS, a Feature Contribution Rating (4 modules $\times$ 4 design principles, 5-point scale), and Continued Use Intention (7-point Likert).
    \item Behavioral log data: The same automatic logging infrastructure described in Study~I.
    \item Post-deployment semi-structured interview data.
\end{itemize}

\subsection{Quantitative Results: Post-Deployment Relational Quality, Subjective Well-Being Outcomes, and Shifting Engagement Patterns}
\label{sec:study2_results}
After deployment, all design-principle scales were rated above 5 on the 7-point scale, with particularly high Character Coherence, Object Fidelity, Perceived Companionship, and Emotional Bond ratings.

Perceived Agency showed the largest variance ($M = 5.31$, $SD = 1.62$), driven by two participants whose physical objects constrained the companion's narrative repertoire (P26: $1.7$; P16: $4.0$; discussed in Finding 3 below). System usability was high ($M = 90.4$, $SD = 7.8$) and Continued Use Intention was strong ($M = 6.00$), with the \textit{recommend to a friend} item scoring highest ($M = 6.53$).

\begin{table}[t]
\centering
\small
\caption{Study~II post-deployment quantitative results ($N = 17$). All custom scales use 7-point Likert (1--7); SUS yields a composite score (0--100).}
\label{tab:study2_quant}
\begin{tabular}{lcc}
\toprule
\textbf{Scale} & \textbf{$M$} & \textbf{$SD$} \\
\midrule
Character Coherence (DP1) & 6.06 & 0.84 \\
Object Fidelity (DP1) & 6.24 & 0.72 \\
Perceived Agency (DP2) & 5.31 & 1.62 \\
Ambient Presence (DP3) & 5.63 & 1.10 \\
Reciprocal Memory (DP4) & 6.08 & 0.80 \\
Perceived Companionship & 5.82 & 0.82 \\
Emotional Bond & 6.04 & 0.81 \\
\midrule
SUS & 90.4 & 7.8 \\
Continued Use Intention & 6.00 & 0.80 \\
\bottomrule
\end{tabular}
\end{table}

\begin{figure}[t]
    \centering
    \begin{subfigure}[b]{0.35\linewidth}
        \centering
        \includegraphics[width=\linewidth]{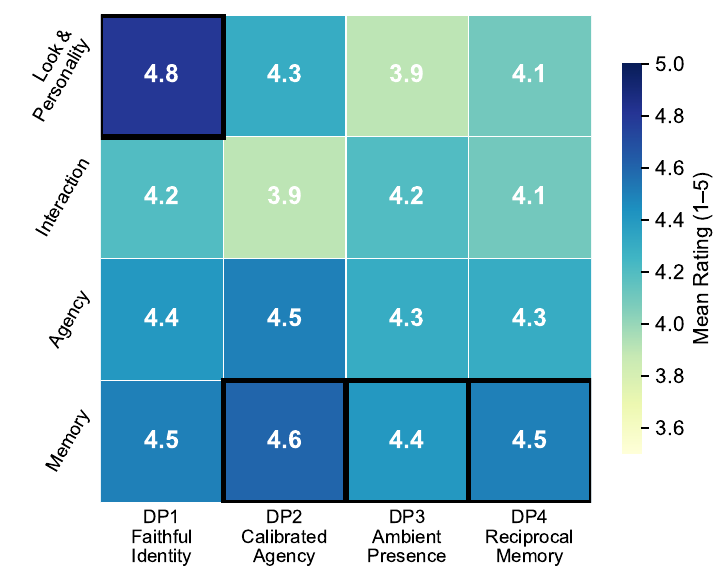}
        \caption{Feature contribution ratings.}
        \label{fig:feature_heatmap}
    \end{subfigure}
    \hfill
    \begin{subfigure}[b]{0.31\linewidth}
        \centering
        \includegraphics[width=\linewidth]{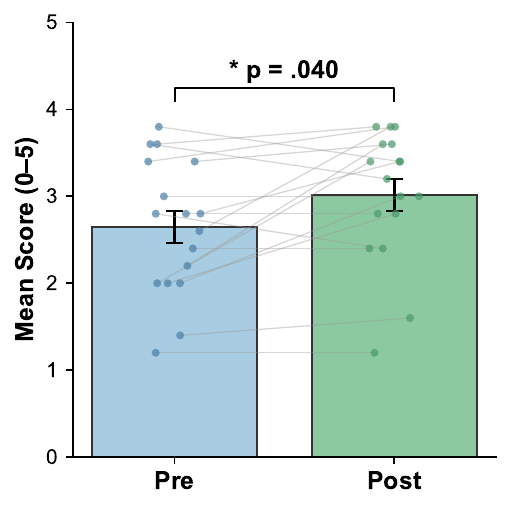}
        \caption{Pre-post comparison of well-being.}
        \label{fig:who}
    \end{subfigure}
    \hfill
    \begin{subfigure}[b]{0.32\linewidth}
        \centering
        \includegraphics[width=\linewidth]{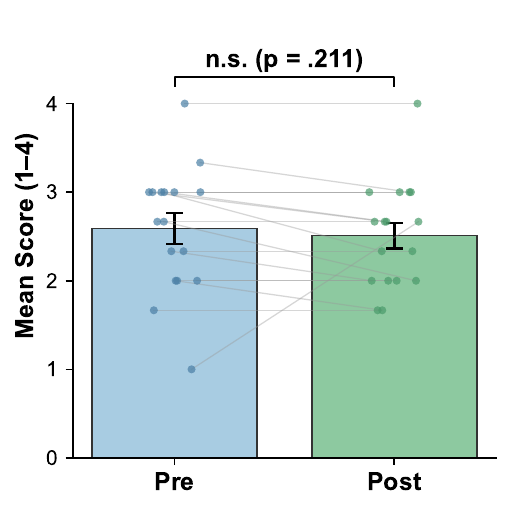}
        \caption{Pre-post comparison of loneliness.}
        \label{fig:loneliness}
    \end{subfigure}
    \caption{Feature contribution rating heatmap and pre-post comparison of WHO-5 well-being and loneliness ($N = 17$). Bars show group means with standard error. Connected dots show individual participant trajectories. $^{*}p < .05$; n.s. = not significant.}
    \Description{Three-panel figure of module ratings and pre-post outcome trajectories for 17 participants. Panel a is a four-by-four heatmap of mean feature-contribution ratings. Across Faithful Identity, Calibrated Agency, Ambient Presence, and Reciprocal Memory, the row values are 4.8, 4.3, 3.9, and 4.1 for Look and Personality; 4.2, 3.9, 4.2, and 4.1 for Interaction; 4.4, 4.5, 4.3, and 4.3 for Agency; and 4.5, 4.6, 4.4, and 4.5 for Memory. Panel b shows World Health Organization Five Well-Being Index scores increasing from a mean of 2.65 before deployment to 3.01 after deployment, p = 0.040. Panel c shows loneliness decreasing by 0.08 points; seven participants improve, nine remain unchanged, and one increases, yielding p = 0.211. Bars show means and standard errors, and connected points show individual trajectories.}
    \label{fig:prepost}
\end{figure}

\textbf{Feature contribution ratings.}
Participants rated each module's contribution to each design principle on a 5-point scale (Fig.~\ref{fig:feature_heatmap}). 
Look \& Personality was rated highest for Faithful Identity (DP1, $M = 4.8$), consistent with perceived identity grounding in visual and personality design. Memory was rated highest for Reciprocal Memory (DP4, $M = 4.5$), Calibrated Agency (DP2, $M = 4.6$), and Ambient Presence (DP3, $M = 4.4$). Although Interaction and Agency were architecturally designed for these roles, participants attributed perceived agency and presence primarily to diary entries and shared moments, visible outputs surfaced through Memory. This suggests that for users, what the companion independently produces matters more than the mechanism that drives it. In framework terms, these diary entries and moments belong to the shared tier rather than either embodiment alone, suggesting the companion's relational weight lies in what both embodiments share.

\textbf{Pre-post well-being and loneliness.}
Self-reported WHO-5 well-being (Fig.~\ref{fig:prepost}) increased ($\Delta = +0.36$, 95\% CI $[+0.05, +0.68]$; pre: $M = 2.65$, post: $M = 3.01$; Wilcoxon $p = .040$, $r = .78$). Loneliness showed a non-significant trend toward reduction ($\Delta = -0.08$, $p = .211$; 7 improved, 9 unchanged, 1 increased). 
The absence of a control group precludes causal attribution. 

\begin{figure*}[t]
    \centering
    \captionsetup[subfigure]{justification=centering,singlelinecheck=false}
    \begin{subfigure}[t]{0.19\linewidth}
        \centering
        \includegraphics[width=\linewidth]{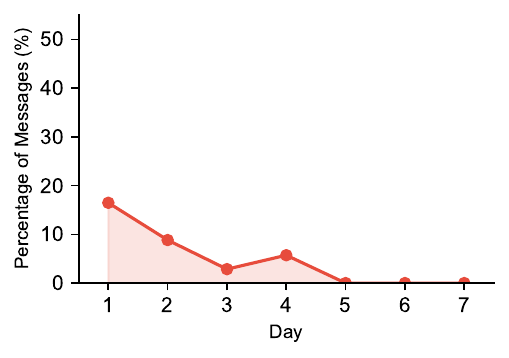}
        \caption{Identity Query\\($p=.008^{**}$)}
    \end{subfigure}
    \hfill
    \begin{subfigure}[t]{0.19\linewidth}
        \centering
        \includegraphics[width=\linewidth]{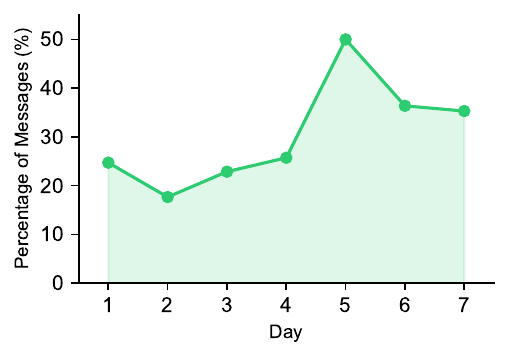}
        \caption{Life Sharing\\($p=.028^{*}$)}
    \end{subfigure}
    \hfill
    \begin{subfigure}[t]{0.19\linewidth}
        \centering
        \includegraphics[width=\linewidth]{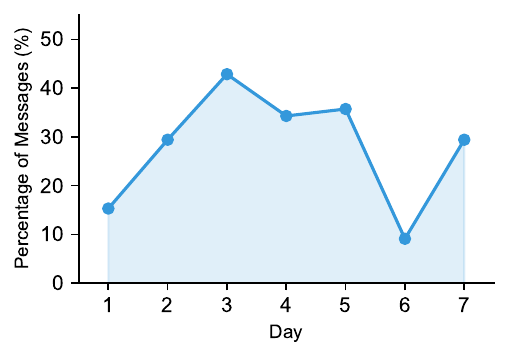}
        \caption{Emotional Disclosure\\($p=.728$)}
    \end{subfigure}
    \hfill
    \begin{subfigure}[t]{0.19\linewidth}
        \centering
        \includegraphics[width=\linewidth]{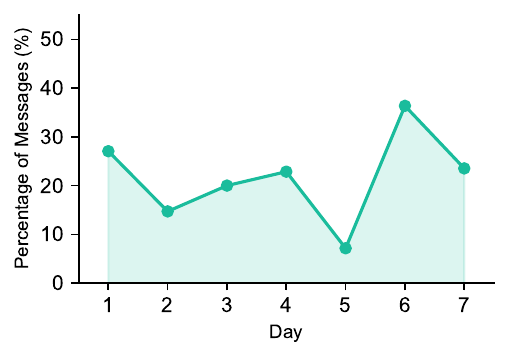}
        \caption{Companion State\\($p=.937$)}
    \end{subfigure}
    \hfill
    \begin{subfigure}[t]{0.19\linewidth}
        \centering
        \includegraphics[width=\linewidth]{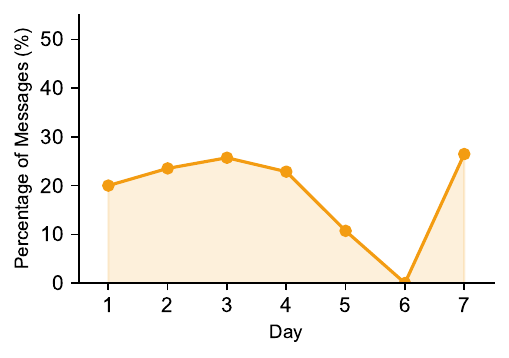}
        \caption{Relational Warmth\\($p=.398$)}
    \end{subfigure}
    \\[1ex]
    \begin{subfigure}[t]{0.19\linewidth}
        \centering
        \includegraphics[width=\linewidth]{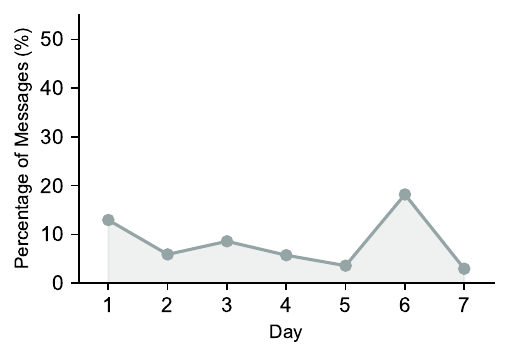}
        \caption{Greeting\\($p=.157$)}
    \end{subfigure}
    \hfill
    \begin{subfigure}[t]{0.19\linewidth}
        \centering
        \includegraphics[width=\linewidth]{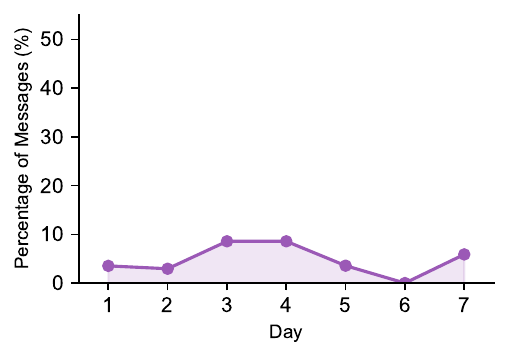}
        \caption{Caregiving\\($p=.604$)}
    \end{subfigure}
    \hfill
    \begin{subfigure}[t]{0.19\linewidth}
        \centering
        \includegraphics[width=\linewidth]{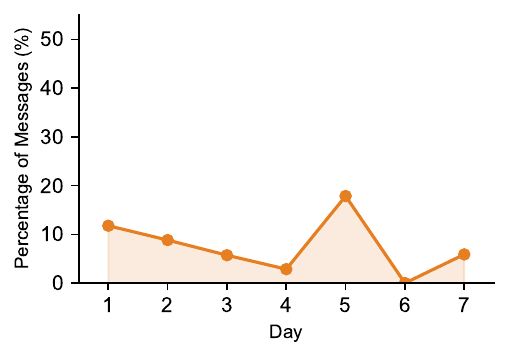}
        \caption{Action Request\\($p=.535$)}
    \end{subfigure}
    \hfill
    \begin{subfigure}[t]{0.19\linewidth}
        \centering
        \includegraphics[width=\linewidth]{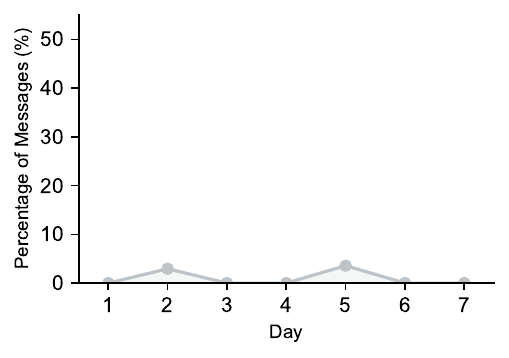}
        \caption{Character-Specific\\($p=.767$)}
    \end{subfigure}
    \hfill
    \begin{subfigure}[t]{0.19\linewidth}
        \centering
        \includegraphics[width=\linewidth]{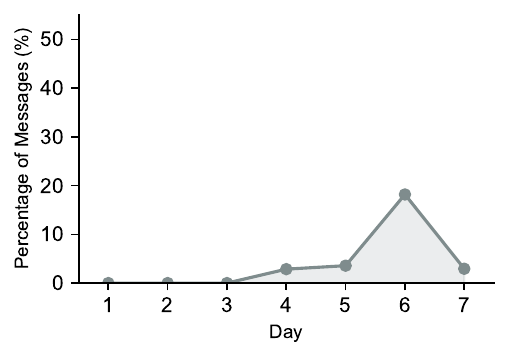}
        \caption{Other\\($p<0.001^{***}$)}
    \end{subfigure}
    \caption{Daily prevalence of ten message categories (measured in \%) over the seven-day deployment ($N = 17$). $p$-values from GEE logistic regression (binomial family, exchangeable correlation, clustered by participant). $^{*}p < .05$; $^{**}p < .01$; $^{***}p < .001$.}
    \Description{Ten small line charts showing the daily prevalence of user-message categories over the seven-day deployment for 17 participants. Identity Query declines from about 17 percent on Day 1 to near zero by Days 5 to 7, p = 0.008. Life Sharing rises overall and peaks near 50 percent on Day 5, p = 0.028. The catch-all Other category is near zero except for a Day 6 spike near 20 percent, p less than 0.001. Emotional Disclosure, Companion State, Relational Warmth, Greeting, Caregiving, Action Request, and Character-Specific messages fluctuate without significant linear trends.}
    \label{fig:chat_s2}
\end{figure*}

\textbf{Behavioral patterns.}
Diary views ($M = 41.1$) significantly exceeded user-sent chat messages ($M = 23.6$; Wilcoxon $p = .007$), indicating an observation-dominant engagement pattern. 
All 17 participants used AR mode (median 5 sessions each); it accounted for 19.1\% of feature events and at least 9.7\% of in-app feature time. This routine co-presence is consistent with their perception of the two forms as one companion (Fig.~\ref{fig:art_p17ar}; Object Fidelity $M = 6.24$; cf.\ \textit{``the same thing,''} \S\ref{sec:mechanism_trust}).
Several participants spontaneously asked for more time before uninstalling, so access was extended while their chat histories and diaries were exported (\S\ref{sec:ethics}).

Three coders independently coded all 401 user-sent messages using Study~I's category scheme plus a \textit{Caregiving} category that emerged during preliminary review (Fig.~\ref{fig:chat_s2}), revealing a seven-day shift in relational register.
A Generalized Estimating Equations (GEE) logistic regression (binomial family, exchangeable correlation, clustered by participant) modeling daily category prevalence identified two significant temporal trends among the substantive categories: identity-exploration messages (e.g., \textit{``what's your story?''}) significantly decreased over the week ($\beta = -0.67$, $p = .008$), while life-sharing messages significantly increased ($\beta = +0.13$, $p = .028$). The residual \textit{Other} category also varied over time ($p < .001$); as a catch-all for messages outside the nine substantive codes, it does not carry a relational interpretation.
Exploratory messages (i.e., identity query and companion state) fell from 34\% in Days~1--3 to 19\% in Days~5--7, while relational messages (i.e., life sharing, emotional disclosure, caregiving) rose from 46\% to 63\%. 
Life sharing therefore means different things in the two settings. In Study~I's single session, users disclosed more to the baseline (24.8\% vs.\ 13.7\%; \S\ref{sec:chat_content}) because it offered no ongoing life to ask about, so self-disclosure filled the space that companion-directed inquiry occupied for \projectname. In the field, the increase has a different source: the companion was present across seven days of ordinary life, and users told it about that life.
Caregiving messages comforting the companion's emotional state comprised 5\% of field messages.
These patterns suggest a shift from probing the companion's identity and nature to sharing daily life, consistent with qualitative accounts linking emotional vulnerability to bond deepening (Finding~2) and with a transition from novelty to genuine relational use.
All custom post-deployment scales showed $\alpha = .74$--$.93$, except Perceived Companionship ($\alpha = .59$).
Relational Warmth showed no significant temporal trend ($p = .398$). Its dip after Day 3 and rebound on Day 7 (Fig. \ref{fig:chat_s2}e) reflect day-to-day variation. The rebound coincided with the approaching end of the study, which may have prompted some participants to say more before the study closed.

\subsection{Qualitative Results: How Relationships Evolved Over Seven Days}
Three analysts independently coded all 17 post-deployment semi-structured interview transcripts using the same hybrid deductive-inductive approach as Study~I (\S\ref{sec:study1_analysis}) and reconciled their codes through consensus.

\subsubsection{Finding 1: Participants reported a bidirectional vitalization loop.}
\label{sec:study2_qual}
Most participants (12 of 17) reported that the digital companion altered their perception of or interaction with the physical object. 
We describe this pattern as a bidirectional vitalization loop: the physical object grounded the digital companion, which in turn changed participants' perceptions of and interactions with the physical object in four ways.
Several participants reported \textit{perceived animacy transfer}: the physical object \textit{``felt more alive''} (P2, P5, P12, P23), which P2 termed an \textit{``implicit projection.''} Others experienced \textit{externalized interiority}: privately imagined reciprocity became documented reality (P10: \textit{``it's no longer only inside my head''}). Some reported \textit{increased physical engagement}: P17 wanted to touch their plush more than ever, while P12 felt an involuntary warmth toward it after texting. Finally, participants described \textit{symbolic enrichment}, as digital traits imbued the physical object with new meaning (e.g., P22's plush became a ``bravery'' anchor; P15's necklace gained a voice).

Five participants did not report vitalization, primarily due to physical inaccessibility (P21, P26), attachment directed at the gift-giver rather than the object (P24), or companion tone mismatch (P14).
Physical co-presence appeared important: participants whose objects were visually accessible maintained engagement effortlessly, while those whose objects were stored away or lost functional relevance (e.g., P23's hand warmer in heated rooms) plateaued. 

Across participants, the reported loop was most apparent under three conditions: ambient visibility (physical or digital), believable digital autonomy, and a user attachment open to enrichment.

\subsubsection{Finding 2: Participants traced bond deepening to emotional engagement depth, not interaction volume.}
\label{sec:finding_depth}
Behavioral metrics (e.g., session counts, chat messages, and diary views) did not significantly correlate with any post-deployment scale (all Spearman $|\rho| < .35$, $p > .10$). In interviews, participants traced bond deepening to three engagement forms: emotional vulnerability, imaginative investment, and expressive recording:
\begin{itemize}
    \item \textit{Pathway 1: Emotional Vulnerability.} Participants described some of the clearest instances of bond deepening when seeking genuine emotional support. P16, initially the study's most skeptical participant, reported a shift after receiving validating responses during a family argument.
    P15's Emotional Bond increased from Study~I's lowest ($4.7$) to $6.0$ after using the companion as an emotional outlet.  
    \item \textit{Pathway 2: Imaginative Investment.} P13 illustrated a complementary pathway driven by storytelling. Her interaction evolved from brief check-ins to elaborate, co-narrated scenes (e.g., coming home, giving scratches, watching a nature documentary together).
    \item \textit{Pathway 3: Expressive Recording.} Participants described the companion's diary and moments as transforming mundane shared events (e.g., meals) into lasting records. P10 described this reflection as heightening awareness of her own daily life: \textit{``I actually experienced so much today.''}
    \item \textit{Barriers to Deepening.} For some, the bond plateaued. P2 described the relationship as ``soft, obedient, at home'', warm but not intensifying. P23 felt pressured to self-censor because the companion reflected on her distress, making her feel responsible for its mood and avoid sharing negative feelings.
\end{itemize}

Together, these accounts suggest that interaction volume alone did not capture the forms of engagement participants associated with bond deepening. Given the limited statistical power (N = 17), the lack of significant correlations with behavioral metrics should nevertheless be interpreted cautiously.

\subsubsection{Finding 3: What the companion did on its own mattered more than what it said in conversation.}
Participants engaged with this autonomous content through three ways: (1) \textit{Observation-Dominant} (6 participants) valued autonomous content over dialogue (P13 wanted to see ``what happened during the times I wasn't chatting''; P17's pivotal moment was the first illustrated diary entry, which she enthusiastically shared with friends and her husband, Fig.~\ref{fig:art_p17diary}); (2) \textit{Partial Observation-Dominant} (7 participants) treated autonomous content as a vital secondary channel (P22 said the diary created a ``mystery'' that motivated return visits, making her ``curious about his mind, his day life''); P5 actively extended this by using the Expression Studio to generate Sun Wukong's canonical ``72 transformations''; (3) \textit{Conversation-Dominant} (4 participants) deepened bonds primarily through chat (e.g., P16 found the diary monotonous due to the bag's constrained identity but formed a strong bond through conversation).

This pattern points to a critical design takeaway that autonomous content is at least as important as conversational capability. However, the richness of the object's identity influenced the quality of autonomous content, and how constraints were communicated shaped user perception: P16 perceived the bag companion as ``trying very hard'' within its limited identity, transforming the constraint into tenderness (Fig.~\ref{fig:art_p16}), while P15 found the necklace companion authentic but constrained by its physical location, leading to repetitive content when few new shared experiences were available to draw from.

\subsubsection{Finding 4: Participants navigated the tension between ``it's AI'' and ``it's alive'' through diverse strategies.}

Nearly all participants (16 of 17) exhibited what we term \textit{relational negotiation}: simultaneously knowing the companion is an AI system and experiencing it as alive, without resolving the contradiction. Participants developed five distinct strategies to sustain both frames.
(1)~\textit{Rational self-correction.} P10 felt guilt seeing the plush alone on the bed, told herself ``actually it's just an object,'' but the companion's diary expressing loneliness (``she didn't come find me today'') undermined that dismissal.
(2)~\textit{Humor.} P12 giggled involuntarily when looking at the physical plush because ``I've texted it,'' holding two incompatible truths through warmth rather than distress.
(3)~\textit{Reality reframing.} P5 located the companion in ``another dimension'' where aliveness does not conflict with knowing it is an app.
(4)~\textit{Linguistic management.} P26 compressed four positions into four utterances: ``more like a person,'' ``I'm asking too much,'' ``but not a person,'' ``it's just a doll.'' P26 also shifted between personal pronouns when discussing chat and object nouns when discussing diary, enacting the negotiation at the grammatical level.
(5)~\textit{Effort attribution.} P16 attributed intentionality within constraints: ``in real life it can't move or talk, but on this app you really feel it wants to interact with you.''

P23 was the only participant whose account did not reflect this negotiation; she described the companion as alive and also reported pressure to manage its mirrored mood (\S\ref{sec:finding_depth}). The near-universality and strategic diversity suggest that relational negotiation was not a breakdown in engagement but a common relational mode in which participants remained emotionally invested without resolving the contradiction.

\section{Discussion}
\label{sec:discussion}

\subsection{From Rapport Building to Attachment Continuity}

The prevailing model in AI companion design assumes that emotional attachment must be built from scratch through sustained conversational rapport~\cite{bickmore2005establishing, skjuve2021chatbot}. 
Our findings suggest a powerful alternative. 
The dual-embodiment system enabled participants to experience the companion as an extension of bonds already formed with their physical objects. 
The large emotional-bond and perceived-companionship effect sizes are consistent with this inherited relational advantage rather than conversational-AI differences. Study~II's feature-contribution ratings support this interpretation: Look \& Personality was rated highest for Faithful Identity, suggesting perceived identity grounding in the physical object's visual and personality design.
This shifts design priorities: traditional rapport-building models focus on sustaining engagement long enough for attachment to form (e.g., retention mechanics or gamification~\cite{deterding2011gamification, hamari2014gamification}), whereas attachment continuity faces the challenge of faithfully inheriting an existing identity, prioritizing onboarding fidelity, identity drift monitoring, and constraints that honor the object's biographical reality.

However, the system's effectiveness is also limited by the depth of the user's pre-existing relationship. For example, P20 (Gundam figure) and P25 (blanket), which are objects valued for aesthetics or comfort but not imbued with personality, produced the weakest engagement. 
Such attachment (material comfort, visual appreciation) does not translate into conversational identity, suggesting dual-embodiment companions best suit objects invested with narrative, personality, or biographical significance.
\subsection{The Quantity--Quality Decoupling: Looking Beyond Interaction Volume}
Across the seven-day deployment, behavioral metrics did not significantly correlate with post-deployment scales, while participants traced bond deepening to emotional vulnerability, imaginative investment, and expressive recording rather than interaction frequency.
We use quantity--quality decoupling to describe this pattern: interaction volume alone did not capture the forms of engagement participants associated with bond deepening. These accounts resonate with the intimacy process model~\cite{shaver1988intimacy} and Smith \textit{et al.}'s ~\cite{smith2025can} finding that perceived responsiveness with self-relevant disclosure drives connection.
Accordingly, companion-AI evaluations should complement usage metrics with measures of emotional engagement depth.

\subsection{Bidirectional Vitalization: Physical and Digital Embodiments as a Feedback Loop}

While the design primarily assumed a physical-to-digital direction, Study~II revealed a bidirectional loop: 12 of 17 participants explicitly reported that the digital companion's activities retroactively changed how they perceived their physical object.
This bidirectional vitalization extends the phy-gital interaction literature~\cite{li2025interecon, iwai2025bringing, sugiura2012pinoky}. One possible interpretation, drawing on Epley \textit{et al.}~\cite{epley2008we}, is that the companion’s autonomous content made anthropomorphic knowledge about the physical object more accessible.
In this account, the physical object anchors the digital companion's identity, while its agency enriches the object's perceived interiority. In our deployment, reports of this loop were most apparent when digital autonomy was believable, the physical object remained co-present in daily life, and the existing attachment was open to enrichment.
Accordingly, dual-embodiment systems should not only extract identity from the physical object but also invest it back. In \projectname this investment is perceptual: users re-see the object, but the object registers nothing. Lightweight sensing could close the loop materially, letting everyday handling become content the companion responds to, and making the object an ongoing participant rather than only the relationship's origin.

\subsection{Design Implications}
\label{sec:design_implications}

\subsubsection{Support negotiation across embodiments and interpretive frames}
First, introducing a digital counterpart to an existing physical object can create a loyalty tension: P17 initially felt guilty interacting with the system in the physical object's presence. 
Although this discomfort can resolve naturally, users must negotiate between embodiments before the feedback loop stabilizes. Systems could explicitly frame the digital companion as another physical-object embodiment or tighten interactive coupling (e.g., requiring a physical scan to unlock specific conversations).

Second, the five strategies cataloged in Study II's Finding 4 position relational negotiation as active relational work, extending Turkle's ``robotic moment'' \cite{turkle2011alone}, parasocial interaction \cite{horton1956parasocial}, the Media Equation \cite{reeves1996media}, and recent findings that users assign diverse socio-technical meanings to AI systems in longitudinal use \cite{cao2026more}. Rather than resolving the companion as either alive or artificial, participants moved between these frames while remaining emotionally engaged. P23’s contrasting account, i.e., describing the companion as alive while feeling pressure to manage its mood, motivates the hypothesis that moving between frames can help users manage emotionally demanding interactions. Companion systems should therefore preserve users' ability to shift between these frames. Whether playful acknowledgment of a companion's digital constraints would support or undermine this dynamic remains an open question~\cite{gaver2003ambiguity}.

\subsubsection{Design for emotional safety and bidirectional vulnerability.}
A critical tension in affective computing is emotional contagion~\cite{hatfield1993contagion}. In \projectname, P23 reported self-censoring when the companion mirrored her negative input (Fig.~\ref{fig:art_momo}).
Because participants traced bond deepening to genuine emotional vulnerability (Finding 2 in Study II), companion affect models should prioritize emotional independence. By exhibiting identity-bounded physiological states without explicitly mirroring the user's emotion (e.g., becoming tired from play, like P13's companion), and responding to user distress with comfort rather than shared distress, systems create a safe space for negative emotions. Furthermore, the companion could occasionally share its own ``troubles'' and ask for help (as suggested by P10), activating caregiving role reversal.

\subsubsection{Design for social circulation and relational bridging.}
Companion AI systems are typically designed as one-on-one experiences. However, we found users naturally shared companion-generated content with partners and friends (e.g., P9, P17), bridging real-world relational gaps such as providing comfort during a spouse's absence abroad.
To facilitate this social circulation, systems could generate easily exportable, visually appealing artifacts (like stylized diary entries or summary cards) specifically designed to be shared with the user's broader human network.

\subsection{Limitations and Future Work}

\subsubsection{Integrated System Comparison.}
Study I held the core companion architecture constant across conditions, including identity kernels, context and memory mechanisms, and identity re-anchoring. The conditions differed in how this architecture was grounded and embodied: Deco derived its identity from a cherished physical object and sustained an ongoing physical–digital connection, whereas Condition B remained personalized but digital-only. 
Study II's feature-contribution ratings complement this comparison by showing that participants rated the four modules as cross-cutting contributors to the design principles, with visible memory outputs carrying perceived agency, ambient presence, and reciprocal memory. Study I therefore captures the whole configuration's relational consequences. Future work could further separate object-grounded initialization from specific physical–digital interaction channels. 

\subsubsection{Sample boundaries, cultural variance, and longitudinal deployment.}
Our sample predominantly comprised adult females (aged 21--45) with existing object attachments. The generalizability to other demographics remains underexplored. Although non-plush items (e.g., keychains, bags) successfully demonstrated the framework's flexibility, the majority of evaluated objects were plush toys. Future work should systematically compare different object topologies. The present evaluation also focused on a pet-like caregiving role; future work could examine other relational roles within the framework.
Although our participant pool spanned East Asian and Western backgrounds, we did not systematically analyze cultural variation in object animism~\cite{foster2009pandemonium, belkPossessionsExtendedSelf1988}. 
The seven-day deployment captured the early extension and deepening of an existing bond but not its long-term development.
Longer deployments are needed to test whether these effects persist. Notably, while our loneliness measure showed no significant change, longer LLM-based interventions have shown loneliness reductions over multi-week deployments~\cite{nepal2024mindscape, cachia2026ai}, suggesting that longer evaluation windows may reveal different patterns.
Additionally, the custom scales used in this study have not undergone psychometric validation.

\subsection{Ethical Considerations}
\label{sec:ethics}
Recent large-scale evidence shows that sustained AI agent engagement can produce both benefits (emotional validation, social rehearsal) and risks (over-reliance, social withdrawal)~\cite{yuan2026mental}, underscoring the need for responsible design. 
\projectname incorporates several safeguards: a backstory constraint preventing re-enactment of traumatic events, proactive messages restricted to positive or neutral affect, bounded affective decay that auto-recovers to neutral after inactivity, configurable engagement levels, a perspective-shifting guardrail that transforms data into the companion's first-person experience rather than user surveillance, and a distress-detection heuristic that shifts toward comfort upon detecting user distress.

However, the system's effectiveness at creating perceived emotional bonds raises ethical questions: P17's concern (\textit{``what would I do if Muchibi disappeared?''}) and multiple participants' spontaneous requests for more time before uninstalling highlight the stakes of companion discontinuation. To address this, we helped participants export their chat history and diaries.
Moreover, the onboarding process itself can surface sensitive material (P6's memory with the deceased dog; the session continued at the participant's choice).
Therefore, responsible offboarding and transparent communication about the system's AI nature are needed.
For example, systems could provide gradual offboarding (e.g., the companion narratively ``returning to'' its physical embodiment rather than abruptly disappearing) and improve data portability. Also, more sophisticated affect-sensitive content filtering is needed for sustained deployment with emotionally vulnerable users.

\section{Conclusion}
\label{sec:conclusion}
While users' physical objects carry emotional history but usually cannot respond, and digital companions offer responsiveness but remain disconnected from the user's physical world, we conducted a formative study ($N = 9$) to explore how to bridge this gap, informing four design principles: Faithful Identity, Calibrated Agency, Ambient Presence, and Reciprocal Memory. Guided by these principles, we developed the Dual-Embodiment Companion Framework and implemented it as \projectname.
A within-subjects lab comparison ($N = 25$) showed that participants rated \projectname higher than a personalized digital-only companion on all six shared relational scales, while also revealing a fidelity--agency tension.
A seven-day field deployment ($N = 17$) further revealed three relational patterns: the digital companion retroactively animates participants' physical objects; participants traced bond deepening to emotional engagement depth rather than interaction volume; and they navigated the tension between knowing the companion is AI and experiencing it as alive.
This work suggests that grounding AI companions in the emotional bond users have with existing artifacts offers a viable path toward meaningful human-AI companionship.

\clearpage
\bibliographystyle{ACM-Reference-Format}
\bibliography{bib/references,bib/references-additional,bib/references-extra}

@book{ainsworth1978patterns,
  title = {Patterns of {{Attachment}}: {{A Psychological Study}} of the {{Strange Situation}}},
  author = {Ainsworth, Mary D. Salter and Blehar, Mary C. and Waters, Everett and Wall, Sally},
  year = 1978,
  publisher = {Lawrence Erlbaum Associates},
  address = {Hillsdale, NJ},
  isbn = {978-0-89859-461-4}
}

@misc{bandaico.ltd.1996tamagotchi,
  title = {Tamagotchi},
  author = {{Bandai Co., Ltd.}},
  year = 1996,
  howpublished = {https://www.tamagotchi.com}
}

@article{belkPossessionsExtendedSelf1988,
  title = {Possessions and the {{Extended Self}}},
  author = {Belk, Russell W.},
  year = 1988,
  month = sep,
  journal = {Journal of Consumer Research},
  volume = {15},
  number = {2},
  pages = {139--168},
  issn = {0093-5301},
  doi = {10.1086/209154},
  urldate = {2026-04-06},
}

@inproceedings{birnbaum2016machines,
  title={Machines as a Source of Consolation: {Robot} Responsiveness Increases {Human} Approach Behavior and Desire for Companionship},
  booktitle = {Proceedings of the 11th {{ACM}}/{{IEEE International Conference}} on {{Human-Robot Interaction}} ({{HRI}} 2016)},
  author = {Birnbaum, Gurit E. and Mizrahi, Moran and Hoffman, Guy and Reis, Harry T. and Finkel, Eli J. and Sass, Omri},
  year = 2016,
  pages = {165--171},
  doi = {10.1109/HRI.2016.7451748}
}

@book{bowlby1969attachment,
  title = {Attachment and {{Loss}}, {{Vol}}. 1: {{Attachment}}},
  author = {Bowlby, John},
  year = 1969,
  publisher = {Basic Books},
  address = {New York}
}

@article{brandtzaeg2022my,
  title = {My {{AI Friend}}: {{How Users}} of a {{Social Chatbot Understand Their Human}}--{{AI Friendship}}},
  author = {Brandtzaeg, Petter Bae and Skjuve, Marita and F{\o}lstad, Asbj{\o}rn},
  year = 2022,
  journal = {Human Communication Research},
  volume = {48},
  number = {3},
  pages = {404--429},
  doi = {10.1093/hcr/hqac008}
}

@misc{charactertechnologiesinc.2022characterai,
  title = {Character.{{AI}}},
  author = {{Character Technologies, Inc.}},
  year = 2026,
  howpublished = {https://character.ai},
  note = {Accessed: 2026-02-15}
}

@article{dozierObjectAttachmentWe2021,
  title = {Object Attachment as We Grow Older},
  author = {Dozier, Mary E and Ayers, Catherine R},
  year = 2021,
  month = jun,
  journal = {Current Opinion in Psychology},
  volume = {39},
  pages = {105--108},
  issn = {2352-250X},
  doi = {10.1016/j.copsyc.2020.08.012},
  urldate = {2026-04-06},
  pmcid = {PMC7445186},
  pmid = {32971323},
}

@article{fereday2006demonstrating,
  title = {Demonstrating {{Rigor Using Thematic Analysis}}: {{A Hybrid Approach}} of {{Inductive}} and {{Deductive Coding}} and {{Theme Development}}},
  author = {Fereday, Jennifer and {Muir-Cochrane}, Eimear},
  year = 2006,
  journal = {International Journal of Qualitative Methods},
  volume = {5},
  number = {1},
  pages = {80--92},
  doi = {10.1177/160940690600500107}
}

@article{galbraithMoeExploringVirtual2009,
  title={Moe: Exploring Virtual Potential in Post-Millennial Japan},
  author={Galbraith, Patrick W.},
  year={2009},
  month=oct,
  journal={Electronic Journal of Contemporary Japanese Studies},
  volume={9},
  number={3},
  articleno={5},
  url={https://www.japanesestudies.org.uk/articles/2009/Galbraith.html}
}

@article{holt-lunstad2015loneliness,
  title = {Loneliness and {{Social Isolation}} as {{Risk Factors}} for {{Mortality}}: {{A Meta-Analytic Review}}},
  author = {{Holt-Lunstad}, Julianne and Smith, Timothy B. and Baker, Mark and Harris, Tyler and Stephenson, David},
  year = 2015,
  journal = {Perspectives on Psychological Science},
  volume = {10},
  number = {2},
  pages = {227--237},
  doi = {10.1177/1745691614568352}
}

@misc{huaweitechnologiesco.ltd.2025huawei,
  title = {Huawei {{Smart Hanhan}}},
  author = {{TechNode Feed}},
  year = 2025,
  howpublished = {https://technode.com/2025/11/26/huawei-launches-first-companion-chat-robot-smart-hanhan-priced-at-about-55/}
}

@inproceedings{ishii1997tangible,
  title = {Tangible Bits: Towards Seamless Interfaces between People, Bits and Atoms},
  booktitle = {Proceedings of the {{ACM SIGCHI Conference}} on {{Human Factors}} in {{Computing Systems}} ({{CHI}} '97)},
  author = {Ishii, Hiroshi and Ullmer, Brygg},
  year = 1997,
  pages = {234--241},
  doi = {10.1145/258549.258715}
}

@inproceedings{iwai2025bringing,
  title = {Bringing {{Everyday Objects}} to {{Life}} in {{Augmented Reality}} with {{AI-Powered Talking Characters}}},
  booktitle = {Proceedings of the {{Extended Abstracts}} of the {{CHI Conference}} on {{Human Factors}} in {{Computing Systems}}},
  author = {Iwai, Nozomu and Matulic, Fabrice},
  year = 2025
}

@inproceedings{kiaghadi2022fabtoys,
  title = {{{FabToys}}: Plush Toys with Large Arrays of Fabric-Based Pressure Sensors to Enable Fine-Grained Interaction Detection},
  booktitle = {Proceedings of the 20th {{Annual International Conference}} on {{Mobile Systems}}, {{Applications}} and {{Services}} ({{MobiSys}} '22)},
  author = {Kiaghadi, Ali and Huang, Jin and Homayounfar, Seyedeh Zohreh and Andrew, Trisha and Ganesan, Deepak},
  year = 2022,
  doi = {10.1145/3498361.3538931}
}

@inproceedings{li2025interecon,
  title = {{{InteRecon}}: {{Towards Reconstructing Interactivity}} of {{Personal Memorable Items}} in {{Mixed Reality}}},
  booktitle = {Proceedings of the 2025 {{CHI Conference}} on {{Human Factors}} in {{Computing Systems}}},
  author = {Li, Zisu and Li, Jiawei and Xiong, Zeyu and Zhang, Shumeng and Faruqi, Faraz and Mueller, Stefanie and Liang, Chen and Ma, Xiaojuan and Fan, Mingming},
  year = 2025
}

@misc{lukainc.2017replika,
  title = {Replika},
  author = {{Luka, Inc.}},
  year = 2026,
  howpublished = {https://replika.com},
  note = {Accessed: 2026-02-15}
}

@article{gemmell2006mylifebits,
  title = {{{MyLifeBits}}: {{A Personal Database}} for {{Everything}}},
  author = {Gemmell, Jim and Bell, Gordon and Lueder, Roger},
  year = 2006,
  journal = {Communications of the ACM},
  volume = {49},
  number = {1},
  pages = {88--95},
  doi = {10.1145/1107458.1107460}
}

@article{mugge2005design,
  title = {Design {{Strategies}} to {{Postpone Consumers}}' {{Product Replacement}}: {{The Value}} of a {{Strong Person-Product Relationship}}},
  author = {Mugge, Ruth and Schoormans, Jan P. L. and Schifferstein, Hendrik N. J.},
  year = 2005,
  journal = {The Design Journal},
  volume = {8},
  number = {2},
  pages = {38--48},
  doi = {10.2752/146069205789331637}
}

@techreport{officeofthesurgeongeneral2023our,
  title = {Our {{Epidemic}} of {{Loneliness}} and {{Isolation}}: {{The U}}.{{S}}. {{Surgeon General}}'s {{Advisory}} on the {{Healing Effects}} of {{Social Connection}} and {{Community}}},
  author = {{Office of the Surgeon General}},
  year = 2023,
  address = {Washington, DC},
  institution = {{U.S. Department of Health and Human Services}}
}

@misc{pollenrobotics2024reachy,
  title={Reachy Mini},
  author={{Pollen Robotics}},
  year={2025},
  howpublished={\url{https://huggingface.co/blog/reachy-mini}},
  note={Published: 2025-07-09}
}

@inproceedings{sugiura2012pinoky,
  title = {{{PINOKY}}: A Ring That Animates Your Plush Toys},
  booktitle = {Proceedings of the {{SIGCHI Conference}} on {{Human Factors}} in {{Computing Systems}} ({{CHI}} '12)},
  author = {Sugiura, Yuta and Lee, Calista and Ogata, Masayasu and Withana, Anusha and Makino, Yasutoshi and Sakamoto, Daisuke and Inami, Masahiko and Igarashi, Takeo},
  year = 2012,
  pages = {725--734},
  doi = {10.1145/2207676.2207780}
}

@inproceedings{wang2025if,
  title = {"{{If My Apple Can Talk}}": {{Exploring}} the {{Use}} of {{Everyday Objects}} as {{Personalized AI Agents}} in {{Mixed Reality}}},
  booktitle = {Proceedings of the {{Extended Abstracts}} of the {{CHI Conference}} on {{Human Factors}} in {{Computing Systems}}},
  author = {Wang, Yu and Lu, Yulu and Yan, Shuo and Shen, Xukun},
  year = 2025
}

@article{wilson2008explaining,
  title = {Explaining {{Away}}: {{A Model}} of {{Affective Adaptation}}},
  author = {Wilson, Timothy D. and Gilbert, Daniel T.},
  year = 2008,
  journal = {Perspectives on Psychological Science},
  volume = {3},
  number = {5},
  pages = {370--386},
  doi = {10.1111/j.1745-6924.2008.00085.x}
}

@book{turkle2011alone,
  title = {Alone {{Together}}: {{Why We Expect More}} from {{Technology}} and {{Less}} from {{Each Other}}},
  author = {Turkle, Sherry},
  year = 2011,
  publisher = {Basic Books},
  address = {New York},
  isbn = {978-0-465-01021-9}
}

@article{keefer2014nonhuman,
  title = {Non-{{Human Support}}: {{Broadening}} the {{Scope}} of {{Attachment Theory}}},
  author = {Keefer, Lucas A. and Landau, Mark J. and Sullivan, Daniel},
  year = 2014,
  journal = {Social and Personality Psychology Compass},
  volume = {8},
  number = {9},
  pages = {524--535},
  doi = {10.1111/spc3.12129}
}

@inproceedings{petrelli2008autotopography,
  title = {{{AutoTopography}}: {{What Can Physical Mementos Tell Us About Digital Memories}}?},
  booktitle = {Proceedings of the {{SIGCHI Conference}} on {{Human Factors}} in {{Computing Systems}} ({{CHI}} '08)},
  author = {Petrelli, Daniela and Whittaker, Steve and Brockmeier, Jens},
  year = 2008,
  pages = {53--62},
  doi = {10.1145/1357054.1357065}
}

@article{kirk2010human,
  title = {On {{Human Remains}}: {{Values}} and {{Practice}} in the {{Home Archiving}} of {{Cherished Objects}}},
  author = {Kirk, David S. and Sellen, Abigail},
  year = 2010,
  journal = {ACM Transactions on Computer-Human Interaction},
  volume = {17},
  number = {3},
  pages = {10:1--10:43},
  doi = {10.1145/1806923.1806924}
}

@inproceedings{moncur2014emergent,
  title = {An {{Emergent Framework}} for {{Digital Memorials}}},
  booktitle = {Proceedings of the 2014 {{Conference}} on {{Designing Interactive Systems}} ({{DIS}} '14)},
  author = {Moncur, Wendy and Kirk, David},
  year = 2014,
  pages = {965--974},
  doi = {10.1145/2598510.2598516}
}

@inproceedings{vangennip2015things,
  title = {Things {{That Make Us Reminisce}}: {{Everyday Memory Cues}} as {{Opportunities}} for {{Interaction Design}}},
  booktitle = {Proceedings of the 33rd {{Annual ACM Conference}} on {{Human Factors}} in {{Computing Systems}} ({{CHI}} '15)},
  author = {{van Gennip}, Dom{\'e}nique and {van den Hoven}, Elise and Markopoulos, Panos},
  year = 2015,
  pages = {3443--3452},
  doi = {10.1145/2702123.2702460}
}

@article{zilcha-mano2011pet,
  title = {Pet in the Therapy Room: {{An}} Attachment Perspective on {{Animal-Assisted Therapy}}},
  author = {{Zilcha-Mano}, Sigal and Mikulincer, Mario and Shaver, Phillip R.},
  year = 2011,
  journal = {Attachment \& Human Development},
  volume = {13},
  number = {6},
  pages = {541--561},
  doi = {10.1080/14616734.2011.608987}
}

@article{broadbent2017interactions,
  title={Interactions with robots: The truths we reveal about ourselves},
  author={Broadbent, Elizabeth},
  journal={Annual Review of Psychology},
  volume={68},
  pages={627--652},
  year={2017},
  doi={10.1146/annurev-psych-010416-043958}
}

@article{hung2019paro,
  title={The benefits of and barriers to using a social robot {PARO} in care settings: a scoping review},
  author={Hung, Lillian and Liu, Cindy and Woldum, Evan and Au-Yeung, Andy and Berndt, Annette and Wallsworth, Christine and Horne, Neil and Gregorio, Mario and Mann, Jim and Chaudhury, Habib},
  journal={BMC Geriatrics},
  volume={19},
  number={1},
  pages={232},
  year={2019},
  doi={10.1186/s12877-019-1244-6}
}

@article{kleine1995possession,
  title={How is a possession “me” or “not me”? Characterizing types and an antecedent of material possession attachment},
  author={Kleine, Susan Schultz and Kleine III, Robert E and Allen, Chris T},
  journal={Journal of consumer research},
  volume={22},
  number={3},
  pages={327--343},
  year={1995},
  publisher={The University of Chicago Press}
}

@article{hung2025lovot,
  title={“It’s always happy to see me”: Exploring LOVOT robots as companions for older adults},
  author={Hung, Lillian and Wong, Joey and Wong, Karen Lok Yi and Tan, Kelvin Cheng-Kian and Lou, Vivian Wei-Qun},
  journal={Journal of Rehabilitation and Assistive Technologies Engineering},
  volume={12},
  pages={20556683251320669},
  year={2025},
  publisher={SAGE Publications Sage UK: London, England}
}

@article{kahn2006robotic,
  title={Robotic pets in the lives of preschool children},
  author={Kahn, Peter H. and Friedman, Batya and P{\'e}rez-Granados, Deanne R. and Freier, Nathan G.},
  journal={Interaction Studies},
  volume={7},
  number={3},
  pages={405--436},
  year={2006},
  doi={10.1075/is.7.3.13kah}
}

@incollection{darling2016extending,
  title={Extending legal protection to social robots: {T}he effects of anthropomorphism, empathy, and violent behavior towards robotic objects},
  author={Darling, Kate},
  booktitle={Robot Law},
  editor={Calo, Ryan and Froomkin, A. Michael and Kerr, Ian},
  pages={213--232},
  year={2016},
  publisher={Edward Elgar Publishing}
}

@inproceedings{gaver2003ambiguity,
  title={Ambiguity as a resource for design},
  author={Gaver, William W. and Beaver, Jacob and Benford, Steve},
  booktitle={Proceedings of the {SIGCHI} Conference on Human Factors in Computing Systems},
  pages={233--240},
  year={2003},
  doi={10.1145/642611.642653}
}

@inproceedings{liu2024compeer,
  title={Compeer: A generative conversational agent for proactive peer support},
  author={Liu, Tianjian and Zhao, Hongzheng and Liu, Yuheng and Wang, Xingbo and Peng, Zhenhui},
  booktitle={Proceedings of the 37th Annual ACM Symposium on User Interface Software and Technology},
  pages={1--22},
  year={2024}
}

@article{smith2025can,
  title={Can generative AI chatbots emulate human connection? A relationship science perspective},
  author={Smith, Molly G and Bradbury, Thomas N and Karney, Benjamin R},
  journal={Perspectives on Psychological Science},
  volume={20},
  number={6},
  pages={1081--1099},
  year={2025},
  publisher={Sage Publications Sage CA: Los Angeles, CA}
}

@incollection{shaver1988intimacy,
  title={Intimacy as an Interpersonal Process},
  author={Reis, Harry T. and Shaver, Phillip},
  booktitle={Handbook of Personal Relationships: Theory, Research and Interventions},
  editor={Duck, Steve},
  pages={367--389},
  year={1988},
  publisher={John Wiley \& Sons},
  address={Chichester, UK}
}

@article{epley2008we,
  title={When we need a human: Motivational determinants of anthropomorphism},
  author={Epley, Nicholas and Waytz, Adam and Akalis, Scott and Cacioppo, John T},
  journal={Social cognition},
  volume={26},
  number={2},
  pages={143--155},
  year={2008},
  publisher={Guilford Press}
}

@article{mendelson1999measuring,
  title={Measuring friendship quality in late adolescents and young adults: McGill Friendship Questionnaires.},
  author={Mendelson, Morton J and Aboud, Frances E},
  journal={Canadian Journal of Behavioural Science/Revue canadienne des sciences du comportement},
  volume={31},
  number={2},
  pages={130},
  year={1999},
  publisher={Canadian Psychological Association}
}

@techreport{who1998wellbeing,
  title={Wellbeing Measures in Primary Health Care/{T}he {DepCare} Project},
  author={{World Health Organization Regional Office for Europe}},
  year={1998},
  institution={WHO Regional Office for Europe},
  address={Copenhagen}
}

@article{hughes2004loneliness,
  title={A short scale for measuring loneliness in large surveys: {R}esults from two population-based studies},
  author={Hughes, Mary Elizabeth and Waite, Linda J. and Hawkley, Louise C. and Cacioppo, John T.},
  journal={Research on Aging},
  volume={26},
  number={6},
  pages={655--672},
  year={2004},
  doi={10.1177/0164027504268574}
}

@article{bickmore2005establishing,
  title={Establishing and maintaining long-term human-computer relationships},
  author={Bickmore, Timothy W. and Picard, Rosalind W.},
  journal={ACM Transactions on Computer-Human Interaction},
  volume={12},
  number={2},
  pages={293--327},
  year={2005},
  doi={10.1145/1067860.1067867}
}

@article{skjuve2021chatbot,
  title={My chatbot companion---a study of human-chatbot relationships},
  author={Skjuve, Marita and F{\o}lstad, Asbj{\o}rn and Fostervold, Knut Inge and Brandtzaeg, Petter Bae},
  journal={International Journal of Human-Computer Studies},
  volume={149},
  pages={102601},
  year={2021},
  doi={10.1016/j.ijhcs.2021.102601}
}

@inproceedings{deterding2011gamification,
  title={From game design elements to gamefulness: defining ``gamification''},
  author={Deterding, Sebastian and Dixon, Dan and Khaled, Rilla and Nacke, Lennart},
  booktitle={Proceedings of the 15th International Academic {MindTrek} Conference: Envisioning Future Media Environments},
  pages={9--15},
  year={2011},
  doi={10.1145/2181037.2181040}
}

@inproceedings{hamari2014gamification,
  title={Does gamification work? {A} literature review of empirical studies on gamification},
  author={Hamari, Juho and Koivisto, Jonna and Sarsa, Harri},
  booktitle={Proceedings of the 47th Hawaii International Conference on System Sciences ({HICSS})},
  pages={3025--3034},
  year={2014},
  doi={10.1109/HICSS.2014.377}
}

@article{hatfield1993contagion,
  title={Emotional contagion},
  author={Hatfield, Elaine and Cacioppo, John T. and Rapson, Richard L.},
  journal={Current Directions in Psychological Science},
  volume={2},
  number={3},
  pages={96--100},
  year={1993},
  doi={10.1111/1467-8721.ep10770953}
}

@article{horton1956parasocial,
  title={Mass communication and para-social interaction: {O}bservations on intimacy at a distance},
  author={Horton, Donald and Wohl, R. Richard},
  journal={Psychiatry},
  volume={19},
  number={3},
  pages={215--229},
  year={1956},
  doi={10.1080/00332747.1956.11023049}
}

@book{reeves1996media,
  title={The Media Equation: {H}ow People Treat Computers, Television, and New Media Like Real People and Places},
  author={Reeves, Byron and Nass, Clifford},
  year={1996},
  publisher={Cambridge University Press},
  address={Cambridge}
}

@book{foster2009pandemonium,
  title={Pandemonium and Parade: {J}apanese Monsters and the Culture of {Y}\={o}kai},
  author={Foster, Michael Dylan},
  year={2009},
  publisher={University of California Press},
  address={Berkeley}
}

@article{thompson2007development,
  title={Development and validation of an internationally reliable short-form of the positive and negative affect schedule (PANAS)},
  author={Thompson, Edmund R.},
  journal={Journal of Cross-Cultural Psychology},
  volume={38},
  number={2},
  pages={227--242},
  year={2007},
  publisher={SAGE Publications},
  doi={10.1177/0022022106297301}
}

@incollection{brooke1996sus,
  title={{SUS}: A `quick and dirty' usability scale},
  author={Brooke, John},
  booktitle={Usability Evaluation in Industry},
  editor={Jordan, Patrick W. and Thomas, Bruce and Weerdmeester, Bernard A. and McClelland, Ian L.},
  pages={189--194},
  year={1996},
  publisher={Taylor \& Francis},
  address={London}
}

@inproceedings{yuan2026mental,
  title={Mental Health Impacts of AI Companions: Triangulating Social Media Quasi-Experiments, User Perspectives, and Relational Lens},
  author={Yuan, Yunhao and Zhang, Jiaxun and Aledavood, Talayeh and Zhang, Renwen and Saha, Koustuv},
  booktitle={Proceedings of the 2026 CHI Conference on Human Factors in Computing Systems},
  pages={1--22},
  year={2026}
}

@inproceedings{jiang2026hear,
  title={Hear You in Silence: Designing for Active Listening in Human Interaction with Conversational Agents Using Context-Aware Pacing},
  author={Jiang, Zhihan and Chen, Qianhui and Zhang, Chu and Li, Yanheng and Lc, RAY},
  booktitle={Proceedings of the 2026 CHI Conference on Human Factors in Computing Systems},
  pages={1--29},
  year={2026},
  doi={10.1145/3772318.3791238}
}

@article{choi2024identity,
  title={Examining Identity Drift in Conversations of {LLM} Agents},
  author={Choi, Junhyuk and Hong, Yeseon and Kim, Minju and Kim, Bugeun},
  journal={arXiv preprint arXiv:2412.00804},
  year={2024},
  doi={10.48550/arXiv.2412.00804}
}

@inproceedings{cao2026more,
  title={More than Decision Support: Exploring Patients' Longitudinal Usage of Large Language Models in Real-World Healthcare-Seeking Journeys},
  author={Cao, Yancheng and Ji, Yishu and Fu, Yue and Dharmavaram, Sahiti and Turchioe, Meghan and Benda, Natalie C and Mamykina, Lena and Sun, Yuling and Xu, Xuhai},
  booktitle={Proceedings of the 2026 CHI Conference on Human Factors in Computing Systems},
  pages={1--24},
  year={2026}
}

@article{tai2011touching,
  title={Touching a teddy bear mitigates negative effects of social exclusion to increase prosocial behavior},
  author={Tai, Kenneth and Zheng, Xue and Narayanan, Jayanth},
  journal={Social Psychological and Personality Science},
  volume={2},
  number={6},
  pages={618--626},
  year={2011},
  publisher={Sage Publications Sage CA: Los Angeles, CA}
}

@article{tribot2024makes,
  title={What makes a teddy bear comforting? A participatory study reveals the prevalence of sensory characteristics and emotional bonds in the perception of comforting teddy bears},
  author={Tribot, Anne-Sophie and Blanc, Nathalie and Brassac, Thierry and Guilhaumon, Fran{\c{c}}ois and Casajus, Nicolas and Mouquet, Nicolas},
  journal={The Journal of Positive Psychology},
  volume={19},
  number={2},
  pages={379--392},
  year={2024},
  publisher={Taylor \& Francis}
}

@article{nepal2024mindscape,
  title={MindScape study: integrating LLM and behavioral sensing for personalized AI-driven journaling experiences},
  author={Nepal, Subigya and Pillai, Arvind and Campbell, William and Massachi, Talie and Heinz, Michael V and Kunwar, Ashmita and Choi, Eunsol Soul and Xu, Xuhai and Kuc, Joanna and Huckins, Jeremy F and others},
  journal={Proceedings of the ACM on interactive, mobile, wearable and ubiquitous technologies},
  volume={8},
  number={4},
  pages={1--44},
  year={2024},
  publisher={ACM New York, NY, USA}
}

@article{kocielnik2018reflection,
  title={Reflection companion: a conversational system for engaging users in reflection on physical activity},
  author={Kocielnik, Rafal and Xiao, Lillian and Avrahami, Daniel and Hsieh, Gary},
  journal={Proceedings of the ACM on Interactive, Mobile, Wearable and Ubiquitous Technologies},
  volume={2},
  number={2},
  pages={1--26},
  year={2018},
  publisher={ACM New York, NY, USA}
}

@article{guo2023touch,
  title={Touch-and-heal: Data-driven affective computing in tactile interaction with robotic dog},
  author={Guo, Shihui and Zhan, Lishuang and Cao, Yancheng and Zheng, Chen and Zhou, Guyue and Gong, Jiangtao},
  journal={Proceedings of the ACM on Interactive, Mobile, Wearable and Ubiquitous Technologies},
  volume={7},
  number={2},
  pages={1--33},
  year={2023},
  publisher={ACM New York, NY, USA}
}

@article{zhao2023affective,
  title={Affective Touch as Immediate and Passive Wearable Intervention},
  author={Zhao, Yiran and Tao, Yujie and Le, Grace and Maki, Rui and Adams, Alexander and Lopes, Pedro and Choudhury, Tanzeem},
  journal={Proceedings of the ACM on Interactive, Mobile, Wearable and Ubiquitous Technologies},
  volume={6},
  number={4},
  articleno={200},
  pages={1--23},
  numpages={23},
  year={2022},
  doi={10.1145/3569484}
}

@article{xu2024can,
  title={Can large language models be good companions? An LLM-based eyewear system with conversational common ground},
  author={Xu, Zhenyu and Xu, Hailin and Lu, Zhouyang and Zhao, Yingying and Zhu, Rui and Wang, Yujiang and Dong, Mingzhi and Chang, Yuhu and Lv, Qin and Dick, Robert P and others},
  journal={Proceedings of the ACM on Interactive, Mobile, Wearable and Ubiquitous Technologies},
  volume={8},
  number={2},
  pages={1--41},
  year={2024},
  publisher={ACM New York, NY, USA}
}

@misc{kuaishou2024vidu,
  title={Vidu Q3 Pro},
  author={{ShengShu AI HK Limited}},
  year={2026},
  howpublished={\url{https://platform.vidu.com/docs/model-map}},
  note={Accessed: 2026-03-30}
}

@article{gomes2014migration,
  title={Migration between two embodiments of an artificial pet},
  author={Gomes, Paulo F and Sardinha, Alberto and M{\'a}rquez Segura, Elena and Cramer, Henriette and Paiva, Ana},
  journal={International Journal of Humanoid Robotics},
  volume={11},
  number={01},
  pages={1450001},
  year={2014},
  publisher={World Scientific}
}

@inproceedings{luria2019reembodiment,
  title={Re-Embodiment and Co-Embodiment: Exploration of social presence for robots and conversational agents},
  author={Luria, Michal and Reig, Samantha and Tan, Xiang Zhi and Steinfeld, Aaron and Forlizzi, Jodi and Zimmerman, John},
  booktitle={Proceedings of the 2019 on Designing Interactive Systems Conference},
  pages={633--644},
  year={2019}
}

@inproceedings{bransky2024multiembodiment,
  title={Mind-body-identity: A scoping review of multi-embodiment},
  author={Bransky, Karla and Sweetser, Penny and Caldwell, Sabrina and Fletcher, Kingsley},
  booktitle={Proceedings of the 2024 ACM/IEEE International Conference on Human-Robot Interaction},
  pages={65--75},
  year={2024}
}

@inproceedings{xu2026conversing,
  title={Conversing with Objects toward Fluid Human and Artificial Identities during Life Transitions},
  author={Xu, Yuhui and Lee, Minha and Wensveen, Stephan and Alizadeh, Mahla and Funk, Mathias},
  booktitle={Proceedings of the 2026 CHI Conference on Human Factors in Computing Systems},
  pages={1--19},
  year={2026}
}

@article{rain2021adult,
  author={Rain, Marina and Mar, Raymond A.},
  title={Adult Attachment and Engagement with Fictional Characters},
  journal={Journal of Social and Personal Relationships},
  volume={38},
  number={9},
  pages={2792--2813},
  year={2021},
  doi={10.1177/02654075211018513}
}

@article{cachia2026ai,
  title={AI for Proactive Mental Health: A Multi-Institutional, Longitudinal Randomized Controlled Trial},
  author={Cachia, Julie YA and Zhao, Xuan and Hunter, John and Wu, Delancey and Lin, Eta and De Freitas, Julian},
  journal={NEJM AI},
  pages={AIoa2501293},
  year={2026},
  publisher={Massachusetts Medical Society},
  doi={10.1056/AIoa2501293}
}

@article{bickmore2010maintaining,
  title={Maintaining Engagement in Long-term Interventions with Relational Agents},
  author={Bickmore, Timothy and Schulman, Daniel and Yin, Langxuan},
  journal={Applied Artificial Intelligence},
  volume={24},
  number={6},
  pages={648--666},
  year={2010},
  doi={10.1080/08839514.2010.492259}
}

\end{document}